\documentstyle[12pt,aasms4]{article}
\everymath{\rm}
\everydisplay{\rm}
\received{}
\revised{}
\accepted{}
\cpright{}{}\journalid{}{}
\articleid{}{}
\paperid{}
\slugcomment{To appear in {\it The Astrophysical Journal}}
\lefthead{Henry, Kwitter, \& Dufour}
\righthead{The Helix Nebula}
\begin{document}
\title{Morphology and Composition of the Helix Nebula}
\author{R.B.C. Henry\footnote{Visiting Astronomer, Kitt Peak
National Observatory, National Optical Astronomy Observatories,
which is operated by the Association of Universities for Research in
Astronomy, Inc. (AURA) under cooperative agreement with the
National Science Foundation.}}
\affil{Department of Physics \& Astronomy, University of
Oklahoma, Norman, OK  73019; henry@phyast.nhn.ou.edu}
\author{K.B. Kwitter$^1$}
\affil{Department of Astronomy, Williams College, Williamstown,
MA  01267; kkwitter@williams.edu}
\and
\author{R.J. Dufour\footnote{Visiting Astronomer, Palomar
Observatory, operated by the California Institute of 
Technology.}$^,$\footnote{Guest Observer, International Ultraviolet Explorer 
Satellite, which was operated by NASA, ESA, and the SERC.}}
\affil{Department of Space Physics \& Astronomy, Rice University, Houston, TX
77005-1892; rjd@rice.edu}

\begin{abstract}

We present new narrow-band filter imagery in H$\alpha$ and [N~II]
$\lambda$6584 along with UV and optical spectrophotometry measurements
from 1200{\AA} to 9600{\AA} of NGC~7293, the Helix Nebula, a nearby,
photogenic planetary nebula of large diameter and low surface
brightness.  Detailed models of the observable ionized nebula support
the recent claim that the Helix is actually a flattened disk whose
thickness is roughly one-third its diameter with an inner region
containing hot, highly ionized gas which is generally invisible in
narrow-band images.  The outer visible ring structure is of lower
ionization and temperature and is brighter because of a thickening in
the disk.  We also confirm a central star effective temperature and
luminosity of 120,000K and 100L$_{\sun}$, and we estimate a lower limit
to the nebular mass to be 0.30M$_{\sun}$.  Abundance measurements
indicate the following values: He/H=0.12 ($\pm$0.017), O/H=4.60$\times
10^{-4}$ ($\pm$0.18), C/O=0.87 ($\pm$0.12), N/O=0.54 ($\pm$0.14),
Ne/O=0.33 ($\pm$0.04), S/O=3.22$\times 10^{-3}$ ($\pm$0.26), and
Ar/O=6.74$\times 10^{-3}$ ($\pm$0.76). Our carbon abundance
measurements represent the first of their kind for the Helix Nebula.
The S/O ratio which we derive is anomalously low; such values are found
in only a few other planetary nebulae.  The central star properties,
the super-solar values of He/H and N/O, and a solar level of C/O are
consistent with a 6.5M$_{\sun}$ progenitor which underwent three phases
of dredge-up and hot bottom burning before forming the planetary
nebula.

\end{abstract}


\section{Introduction}

The Helix Nebula (NGC~7293) is one of the best known of all planetary
nebulae (PN). Its nearby location (d$\sim$213 pc; Harris et al. 1997)
and large angular size (angular diameter $\sim $960 arcmin) make it
particularly attractive for detailed investigations of PN structure,
morphology and composition. The well known ionized double ring of this
evolved PN is surrounded by a massive molecular envelope which has
also been studied intensively (Healy \& Huggins 1990; Kastner et
al. 1996; Cox et al. 1998). Recent observations with HST have
showcased ``cometary globules" extending radially outward from the
nebular center, which presumably represent ablation from dense clumps
that survived the PN event (O'Dell \& Handron 1996). Observations by
Meaburn et al. (1998) also imply that the likeliest origin of these
knots is the copious and dusty wind from the red giant precursor. 

The central star of the Helix, a DAO white dwarf (Napiwotzki \&
Sch{\"o}nberner 1995), is one of
the hotter and more massive PN central stars.  Recent temperature and mass determinations by G{\'o}rny, Stasi{\'n}ska, \& Tylenda (1997) yield T$_{eff}$ = 117,000K and M$_*$ =
0.93M$_{\odot}$. Abundances in the Helix
have been studied by Hawley (1978), by Peimbert \& Torres-Peimbert
(1987) and by Peimbert, Luridiana, \& Torres-Peimbert (1995). In the
latter two papers it is classified as a Peimbert Type I PN, exhibiting
enhanced N and He abundances. These composition enhancements accord
with predictions that Type~I PN arise from the presumably rare massive
end of the PN progenitor spectrum. In addition, observations of the
molecular envelope indicate significant C~I, implying that the Helix
is also C-rich (Bachiller et al 1997; Young et al. 1997). This
suggests that the third dredge-up has occurred.

To further our understanding of this uniquely accessible object, we
have undertaken a detailed study combining spatially-resolved
spectrophotometry at three locations with narrow-band, flux-calibrated
imaging to create a unified chemical composition model of the
Helix. Our modelling efforts have been guided by O'Dell's (1998)
interpretation that the main body of the Helix is better represented
by a thick disk rather than by a ring.  The model nebulae which we construct are 
constrained by new surface photometry in H$\alpha$ and [N~II] $\lambda$6584.  
Abundance determinations follow the procedure employed in Henry, Kwitter, \& 
Howard (1996) and Kwitter \& Henry (1996; 1998) and
are carried out by combining Final-Archived IUE spectral data for three 
positions in the nebula with ground-based optical spectrophotometry of the same 
positions.

In the next section we present a description of data acquisition and
reduction of the photometric and spectroscopic observations.  In {\S}3
we present our geometrically-tuned models for the Helix as well as the
details of our abundance determinations.  Section~4 contains a
comprehensive discussion of the Helix and speculations about its
progenitor star, while a summary is given in {\S}5.

\section{Observations and Reductions}

\subsection{Imagery}

The imagery observations of NGC 7293 were taken on the night of 1988
December 2 UT
using a focal reducing camera and Tektronix 800$\times$800
CCD on the Palomar 1.5m telescope.  The camera, developed by J.~Hester, covers a 
16-arcminute
field at a resolution of 1.2$''$ per pixel.  The sky was photometric
during the observations and the seeing was estimated to be $\sim$2-3$''$.
Five images were taken through filters isolating H$\alpha$,
[NII]~$\lambda$
6584, and [SII]~$\lambda$6717+31 emission lines, along with line-free
continuum bands (100 \AA~FWHM) around 6450\AA~and 7230\AA.  Details of the
filters and exposure times are given in Table 1A.  In addition to these
primary images,
a series of flat-fields, as well as appropriate exposure dark
frames, were taken
of an illuminated dome screen through each of the five
filters at the beginning and end of the night.

The images were processed in the usual manner for nebular CCD data
using IRAF\footnote{IRAF is distributed by the National Optical Astronomy
Observatories, which is operated by the Association of Universities
for Research in Astronomy, Inc. (AURA) under cooperative agreement
with the National Science Foundation.}.
Each of the five images was dark-subtracted and flat-fielded;
then average sky levels were determined and subtracted.  Photometry of ten
unsaturated stars in the field was done on each of the images and used to
determine scale factors for subtracting the combination of the two
continuum images from each of the three emission line (+continuum) images.
This produced images of the nebula in the light of the emission lines
of H$\alpha$, [NII]~$\lambda$6584, and [SII]~$\lambda$6717+31
only with most stars
and (very weak) nebular continuum removed\footnote{We acknowledge that the
H$\alpha$ image is slightly contaminated by light from [NII]~$\lambda$6548
passing through the blue side transmission wings of the filter, but such
contamination amounts to only a few percent for the worst cases when the
[NII] line is comparable to H$\alpha$ in strength.}.

Figure 1 shows our H$\alpha$ image (without continuum subtraction) with the
aperture locations for the UV and optical spectroscopy discussed
below noted.  The ground-based (KPNO) spectroscopy enabled us to
calibrate the three emission line images to absolute surface brightness in the
lines.  The three KPNO slit locations were mapped onto the continuum-subtracted images and counts through each of the apertures extracted.  These
were compared to the absolute emission line fluxes from the spectra and
calibration constants were derived for each image (in
ergs/cm$^2$/sec/pixel, where each pixel has an area of 1.44
arcsec$^2$).  The
agreement among the three locations between the counts and the
spectral fluxes
was good; the rms errors in the constants for the three positions were 12\%
for H$\alpha$, 4\% for [NII], and 14\% for [SII].

%
%

\subsection{Spectrophotometry}
\subsubsection{UV Data}

The UV spectra were taken with the IUE satellite by R.J.D. during two
observing runs in 1990-1991. UV slit locations and orientations for
positions A, B, and C are shown with ovals in Fig.~1, while the same
information for the optical data discussed below are shown with
rectangles.  Positions A and B were chosen in order to study the
brightened emission in the NE nebular section, while position~C
coincided with the bright SW region.  We obtained SWP and LWP spectra
for positions A, B, and C. The slit position angle was 309$^{\circ}$
for A and C and 322$^{\circ}$ for B.  All spectra are low dispersion
and were taken with the large aperture
(21$\farcs$7$\times$9$\farcs$1).  These data were later reprocessed
as part of the IUE Final Archive\footnote{Spectra in the Final Archive have
been systematically and uniformly re-processed by IUE staff using the
NEWSIPS algorithms, and represent the best available calibration of these
data.} from which our measurements were taken.  Table~1B lists the spectra along with
their integration times.  As an example of the UV data, Figs.~2A,B
display the SWP and LWP spectra of position~B, where some significant
lines are identified. 

\subsubsection{Optical Data}
The optical data were obtained at KPNO during 7-9 December 1996
UT with
the 2.1m telescope plus the Goldcam CCD spectrometer. 
The three slit positions were chosen so as to produce maximum
overlap with the three IUE positions described above.
We employed the Ford 3K $\times$ 1K CCD chip
with 15$\mu$ pixels. We used a 5$\arcsec$- wide slit that extended
285$\arcsec$ in the E-W direction, with a spatial scale of
0$\farcs$78/pixel. Using a combination of two gratings, we obtained
spectral coverage from 3700-9600\AA\ with overlap from
$\sim$5750 - 6750\AA.  For the blue, we used grating \#240 in first
order, blazed at 5500 \AA, with a WG345 blocking filter. Wavelength
dispersion was 1.5 \AA/pixel
($\sim$8 \AA\ FWHM resolution). For the red, grating  \#58, blazed at
8000{\AA} was used in first order with an OG530 blocking filter.  This
yielded 1.9 \AA/pixel ($\sim$10 \AA\ FWHM resolution). The usual bias
and twilight flat-field frames were obtained each night, along with HeNeAr
comparison spectra for wavelength calibration and standard star
spectra for sensitivity calibration.  A summary of the
observational details is presented in Table~1C.

The optical observations were made at three positions previously observed
with the IUE and which are indicated in Fig.~1.
Since the position angle of the Goldcam slit is fixed at
90$\arcdeg$ while the IUE slit position angle is not (cf. Figure 1), the
quality of the overlap varies with IUE position angle; we note also that
because of the  2:1 relative slit widths, the largest possible overlap of the
Goldcam slit onto the IUE slit is $\sim$50\%. Considering slit
position angles and widths of the two data sets, we estimate the
spatial overlap to be around 31\%.

The original images were reduced in the standard fashion using
IRAF. Employing tasks in the {\it kpnoslit\/} package, these two-dimensional
spectra were converted to one dimension by extracting a specific
section along the slit.  
Since the chip is thinned, it produces interference fringes visible in
the red.  In our red spectra the fringes appear at the $\pm$1\% level
at $\sim$7500\AA\ and increase in amplitude with increasing wavelength:
$\pm$1.5\% at 8000\AA, $\pm$4.5\% at 8500\AA, $\pm$6\% at
9000\AA. However, even at their worst, {\it i.e.\/}, at
$\sim$9500{\AA}, the longest wavelength we measure, the fringe
amplitude reaches only about $\pm$7\%. Internal quartz flats
were taken at the position of each object both before and after the
object integrations in anticipation of removing the fringes during
data reduction. As it turned out, however, more noise was introduced
in this process than was removed; we therefore decided to leave the
fringes untouched, and to accept this additional uncertainty in our
line intensities longward of $\sim$7500\AA.  Our reduced optical spectra
for position B are shown in
Figures 2C,D, where again we have marked lines of interest.

\subsubsection{Line Strengths}

Strengths of all optical and UV lines were measured using {\it splot} in IRAF
and are reported in Table~2A.  Fluxes uncorrected for reddening are presented
in columns labelled F($\lambda$), where these flux values have
been normalized to H$\beta$=100 using our observed value of
F$_{H\beta}$ shown in the third row from the bottom of the table.
These line strengths in turn were corrected for reddening by assuming
that the relative strength of H$\alpha$/H$\beta$=2.86 and computing
the logarithmic extinction quantity $c$ shown in the penultimate line of
the table.  Values for the reddening coefficients,
f($\lambda$), are listed in column~(2), where we employed
Seaton's (1979) extinction curve for the UV and that of Savage \&
Mathis (1979) for the optical.

Because of the imperfect spatial overlap between the optical and IUE
observations, a final adjustment was made by multiplying the IUE line
strengths for positions A and B by a merging factor that was determined
from the theoretical ratio of the He~II lines
$\lambda$1640/$\lambda$4686. (Since He~II $\lambda$1640 was unobserved
at position C, no correction was made and we list the merging factor as
unity.)  The calculation of the merging factors is described in detail
in Kwitter \& Henry (1998), and their values are listed in the last row
of Table~2A.

The columns headed I($\lambda$) list our final, corrected line
strengths, again normalized to H$\beta$=100. Uncertainties in line intensities were ascertained by compiling a list of line ratios with values set by atomic constants. Table~2B shows the theoretical values of the line ratios in column~2 followed by observed ratios and percent differences in the next six columns. We note that these line ratios span the optical spectrum. If we allow for the fact that [Ne~III] $\lambda$3968 is problematic because of contamination by H$\epsilon$ and the [S~III] near infrared lines suffer from atmospheric absorption and emission, we conclude that all optical and near infrared line strengths above 10\% of H$\beta$ have uncertainties of $\pm$10\%, lines weaker than that have uncertainties of $\pm$20\%, UV lines are uncertain by $\pm$25\%, and the [S~III] lines in the near infrared are uncertain by more than 50\%. A few line strengths which we feel are particularly problematic are indicated with colons in Table~2A.

\section{Results}

Our two principal goals in this study are to: (1)~use our imagery
of the Helix to examine its gross
physical properties including its morphology; and
(2)~derive accurate chemical abundances
using our UV and optical emission line measurements at three positions 
in the nebula.  We discuss these two analyses in the following
three subsections.

\subsection{Image Analysis - Density and Ionization Variations}

Since the line-of-sight (LOS) reddening to the Helix Nebula is small, the
H$\alpha$ surface brightness at any point, $S_{H\alpha}$, is proportional to
the square of the electron density, $N_e$, along the LOS path through the
ionized medium in that direction: $S_{H\alpha}$ $\propto$ $\int N_e^2~dl$.
While the constant of proportionality involves atomic data and the nebular 
filling factor, in a real nebula
the density varies along the LOS. Nonetheless, for a first approximation, we can 
make the oversimplification that the LOS through each point in the nebula has a
constant density and filling factor, and then use the square root of the 
H$\alpha$
surface brightness as an indication of the spatial {\it variation} in
density across the nebula.  Fig.~3 is a contour map of $\sqrt{S_{H\alpha}}$
normalized to unity in the center of the the nebula.  The contours
show
an elliptical appearance with the highest regions of surface
brightness ``density'' located NNE and SSW of the central star (along PA's of
$\sim$20$^{\circ}$ \& 215$^{\circ}$, respectively).  Relative to the central
``plateau'' of the nebula, the ``density'' enhancements for the NNE and SSW
bright rims are 1.6$\times$ and 1.5$\times$, respectively.  As is evident in Fig.~1, these
bright rims are also where our slit positions B and C are
located.  In addition, a
hand-drawn fit to the contours in our Fig.~3 yields an ellipse
with a major axis along PA=32$\pm$5$^{\circ}$ and an inclination of
28$\pm$10$^{\circ}$.

While developing this paper
we became aware of two notable studies of the Helix Nebula recently
completed.  The first, by Meaburn et al. (1998; hereafter M98),
deals primarily
with the motions and nature of the cometary knots in the central regions of
the nebula.  However, they also present a global model of
the nebula consisting of a
torus (+expanding lobes) containing the bulk of the bright knots, inclined 37$^{\circ}$ with respect to the plane of the sky around an axis
with PA=14$^{\circ}$.  The second, by C. R. O'Dell (1998; hereafter O98),
is a global imagery and long-slit spectroscopy study of the Helix
with similarities to
our project in its spatial coverage and scientific goals. 
However, our two papers differ
in imagery in that our study employs 
the red lines of H$\alpha$, [N~II], and
[S~II], while O98 use the blue-green lines of He~II, H$\beta$, and
[O~III].  Regarding
the true geometrical structure of the Helix Nebula, O98 argues that ``the ring
is actually a disk'' $-$namely a {\it filled} torus inclined
21$^{\circ}$ with respect to the plane of the sky around an axis
oriented with
PA=30$^{\circ}$.

Our H$\alpha$ imagery of the Helix agrees best with the ``filled disk''
model of O98.  This is illustrated in Fig. 4, which shows an inclined
``surface map'' view of the $\sqrt{S_{H\alpha}}$ contour map
shown in Fig. 3 (where north is to the bottom right and east is to the
top right).  In this representation, it is clear that the central part of the
nebula consists of a ``plateau'' in H$\alpha$, rising well
above the sky background.  Figs.~3 and 4 support the filled disk model
of O98, who explained that this morphology is in contrast to
the one usually drawn from ordinary images taken in low-excitation lines such as [N~II] and [S~II], since the inner region is dominated by highly ionized material and thus is not visible.
While this inner plateau is flat to $\pm$10\% or
so in $\sqrt{S_{H\alpha}}$ (density), we also confirm the slight systematic drop in the H$\alpha$ surface brightness at the interface of the inner ring
at distances between 2-3 arcmin from the star.

The signal-to-noise for our calibrated H$\alpha$ and [N~II]~$\lambda$6584 images
is excellent and permits the construction of an [N~II]~$\lambda$6584/H$\alpha$
{\it ratio}
image showing the ionization structure of the nebula in N$^+$ relative to
H$^+$ at a spatial resolution better than 2$''$.  This result is shown in
Fig.~5 for
the full 16$'$ field (left image) 
and for an 8\arcmin{\ } enlarged area around the northern rim (right
image).  Four distinctly different ionization zones in the nebula are
apparent: (a) a nearly circular high ionization inner zone for which the ratio
is 0.35$\pm$0.13; (b) an extended elliptical region with major axis along
PA$\approx$135$^{\circ}$ where generally
the ratio is $\sim$1.3$\pm$0.4, but with many
``plumes'' having lower [N~II]/H$\alpha$ values; (c) ``rims'' in the
NE and SW directions defining the outer boundary of the elliptical region
where the line ratio is highest (2.2-2.6); and 
(d) the outer nebular region which shows a slightly higher ratio than the
central zone but also with larger variations (0.4-0.8) appearing as smoothly-varying wavy structures.

A comparison of Fig.~5 with an LW2 filter ISOCAM map (Cox et
al. 1998; plate~1), where the latter is dominated by molecular hydrogen
emission from dense globules
with neutral cores and ionized surfaces similar to the
photodissociation region morphology seen in
some Galactic H~II regions, strongly suggests that the rims we
see in zone c correspond to material associated with the nebular
ionization front.
However, while the appearance of the NNE and SSW
rims in our [N~II]/H$\alpha$ ratio greyscale map is similar to the LW2 ISOCAM image, many of the structures (separated
clumps) seen in the LW2 image in the SSE part of the nebula are not apparent
in our ratio map.  This suggests that the stellar wind and ionization
parameter are stronger in the SSE (and NNW) parts of the nebula,
producing a thinner ionization front at those locations and correspondingly lower
[N~II] surface brightness on the neutral globules.

Another very apparent, even striking, feature of our [N~II]/H$\alpha$ ratio map
is the appearance of many radial ``plumes'' seen as elongated
structures with lower ratio values crossing the full expanse of the
ring in many places.  Similar structures are evident in the
[O~III]$\lambda$5007/H$\beta$ ratio map of O98.  We interpret these
plumes as extensions of the ``cometary knot'' (CK) structures (O'Dell
\& Handron 1996) seen in the interface between the high ionization
central region and the ring.  O'Dell \& Burkert (1997) present
convincing arguments that the CK are the artifacts of finger-like
density-enhanced structures formed by Rayleigh-Taylor instabilities
arising from the ionization boundary as it progresses outward into the
neutral material ejected from the red giant progenitor of the PN. M98
presented detailed images of the CK in the central regions of the
nebula in H$\alpha$ + [N~II] along with a detailed kinematical model for the
ablated flow of ionized material from the heads of the CK.  Such flows are
slightly supersonic and form a 3D bow shock cylinder around the CK.  Such a
flow around the CK in the central disk of the nebula, or similar neutral dusty
globules in the outer thicker disk, will have enhanced local emission in
[N~II], and [O~III] compared to H~I.  M98 also noted that some of the CK had
definite absorption in their tails, suggesting that the ablated material is
dusty.

The plumes that are observed throughout the thick outer disk of the Helix
in [N~II] (and [O~III]) can thus be explained as enhanced emission from ablated
flows of ionized material from neutral globules in the shell.  The right panel of Figure~5
shows details of these features in the northern part of the ring, which appear
as parabolic ``bow shock''-like structures with bright heads when they are
studied in high contrast [N~II] or [O~III] (O98) images or better, in
[N~II]/H$\alpha$ ratio grey-scaled images where surface brightness effects of
the surrounding nebula are minimized.  The heads of some of these plumes also
appear in Fig 3C of M98 (bottom left part where the ring brightness begins
to grow high), which suggest that they are the same physical phenomena as the
CK, only that they are immmersed in the denser and lower-ionization ring.  While
it is possible that darker insides of the CK could be due to dust obscuration by
elongated neutral globules, it is more likely due to bright rims expected
from looking at a 3D bow shock structure extending nearly along the plane of
the sky.

\subsection{Whole Nebula Model}

While on the sub-arcsecond scale the Helix Nebula is frightfully complex
with density (and possibly temperature) inhomogeneities, our imagery
permits us to further test the proposed geometries of M98 and
O98 by
developing photoionization models of assorted geometries for the
nebula. Thus, we now consider the H$\alpha$ surface brightness
and [N~II]/H$\alpha$ distributions for
the Helix.  To study this, four
radial cuts originating at the central star and extending outward by
600\arcsec~ were made through the H$\alpha$ and [N~II]/H$\alpha$ images shown
in Figs.~1 and 5 for position angles 24\arcdeg, 135\arcdeg, 209\arcdeg, and
315\arcdeg.  The cuts at position
angles 24\arcdeg~ and 209\arcdeg~ were made specifically to intersect
our spectroscopic positions B and C, respectively, and are each
5{\arcsec} in width.  The cuts at 135\arcdeg~ and 315\arcdeg~ are
50\arcsec~
in width and were made
for the purpose of sampling
the smoother ring material away from positions B and C.
The results are presented in
Fig.~6, where the upper panel is a plot of H$\alpha$ surface
brightness in units of
10$^{-15}$ergs/cm$^2$/s/arcsec$^2$
versus distance from the central star in arcseconds for the
cuts at the four position angles.  The lower
panel shows the value of the intensity ratio of [N~II] $\lambda$6584 relative
to H$\alpha$ as a function of distance.  Radial distances
represented in each curve of both panels
have been deprojected according to the position
angle (PA) of the cut by multiplying each radial coordinate
by $\left\{cos^2(PA-22\arcdeg)+
{{sin^2(PA-22\arcdeg)}\over{cos^2(30\arcdeg)}}\right\}^{1/2}$.
We have assumed a
disk inclination with respect to the sky plane of 30\arcdeg~
around an axis whose position angle is 22\arcdeg. (Note that
these angles are roughly averages of values derived by M98 and
O98 quoted in {\S}3.1.)  Each cut is shown with a line of normal width and
types defined in the legend for the specific position angle of
the cut. Also shown with bold solid and dashed lines are model results which will be discussed below.

The position of the brightened ring of the Helix appears clearly between 200\arcsec~
and 300\arcsec~ from the central star, as seen by the broad rise in H$\alpha$
surface brightness for all four position angles at this distance. 
The cuts at 24{\arcdeg} and 209{\arcdeg} show
additional emission due to the presence of the brightened rims
corresponding respectively to our spectroscopic slit
positions B and C (see Fig.~1).

In order to interpret the data in Fig.~6
we have calculated two separate models to approximate the data: one
corresponding to the cuts along position angles 135\arcdeg~ and
315\arcdeg~, which we designate as model 135/315,
and another corresponding to position
angles 24\arcdeg~ and 209\arcdeg, designated as model 24/209.
Notice that the first
directional pair
lie along a line passing through the central star and
extending 600\arcsec~ in each direction, while the second pair
lie nearly along a straight line passing through the central
star roughly orthogonal to the first pair and including
positions B and C.  Both models are
represented in Fig.~6 with bold lines, with the line type
defined in the legend.

The models were
produced in two steps.  First, we calculated a spherical
photoionized model nebula using the code CLOUDY version 90.04 (Ferland 1990),
where the program specifically tabulated
the volume emissivities of H$\alpha$ and [N~II]
$\lambda$6584 as functions of distance from the central star.
We then integrated these
emissivities along lines-of-sight spaced at 10\arcsec~ intervals from
the central star (see Appendix A for detailed specifics of the
line-of-sight calculation) to produce the H$\alpha$ surface
brightnesses
and [N~II]/H$\alpha$ ratios displayed in Fig.~6.  Note that in
converting the linear distance in the photoionization model to
arcseconds we assume a distance to the Helix Nebula of
213~pc (Harris 1997).  Variable parameters in the complete model
included central star temperature and luminosity, and nebular
composition, density, filling factor, and morphology.
In calculating these models we initially 
assumed a spherical nebula for our line-of-sight
calculations but soon found that we could not simultaneously
model the observed H$\alpha$ and [N~II]/H$\alpha$ profiles. 
Thus, we began experimenting with the disk morphology proposed by
O'Dell by
varying the line-of-sight thickness of the nebula as a
function of distance from the central star, assuming that the
nebular disk plane is coincident with the plane of the sky
(recall that the observations have been deprojected).

The final models displayed in Fig.~6 for 135/315 and 24/209 both
employed a blackbody central star effective temperature of 120,000~K,
consistent with the value of 117,000K found by G{\'o}rny,
Stasi{\'n}ska, \& Tylenda (1997) and a luminosity of 100~L$_{\odot}$,
which is similar to the number found by M{\'e}ndez, Kudritzki, \&
Herrero (1992).  The gas phase chemical abundances were set at the
values derived later in {\S}3.3, except in the case of nitrogen, which
was set at 1/3 of its derived value in order to avoid overproducing
[N~II].  This discrepancy is currently unresolvable. Since we believe
that the nitrogen abundances derived at positions A, B, and C below are
reasonably precise, the fault must lie here with our whole-nebular
models. These models apparently have the tendency to produce too much
[N~II] per unit of emission measure, forcing us to reduce the nitrogen
abundance in order to match the surface photometry. Short of suggesting
that all three of the slit positions for our spectroscopy happened to
be coincident with nitrogen-rich knots, it is clear that more detailed
modelling is necessary to understand this discrepancy.
The mass ratio of dust
to gas in the models was 0.018, while the filling factor was 0.55.
The total gas density in the 135/315 model was
60~cm$^{-3}$ throughout.  However, in the case of model 24/209, the total
density was 60~cm$^{-3}$ out to 250\arcsec, then rose to a
maximum of 125~cm$^{-3}$ at 315\arcsec, and finally returned to
60~cm$^{-3}$ at 395\arcsec.  This density
profile was required to match the
H$\alpha$ and [N~II]/H$\alpha$ behavior in Fig.~6.  Additionally,
in the LOS integration of each of the final models, 
the disk thickness (Z) relative to its
diameter (D) was 0.33 from the center out to 200\arcsec, then
gradually increased to 0.66 at 275\arcsec, and returned to 0.33 at
310{\arcsec} in an attempt to imitate O'Dell's model shown in his Fig.~7.

The model results displayed in Fig.~6 provide a reasonable, although
not perfect, fit to the observed profiles.  In the top panel, model
135/315 (bold solid line) is a good fit to the corresponding H$\alpha$
data for all but the outermost region of the nebula. In the bottom
panel we see that the [N~II]/H$\alpha$ prediction for this model is
good out to about 300\arcsec, but rises above the observed line by
roughly 50\% beyond 400\arcsec, primarily because of the predicted
fall-off in H$\alpha$ emission at this location, and suggesting that the
model Str{\"o}mgren edge is actually located at a greater distance.
However, efforts to correct for this, such as altering the gas density
distribution or the stellar luminosity, resulted in an unacceptible
deterioration in the H$\alpha$ predictions in the vicinity of the ring
(200\arcsec-400\arcsec).

Turning to model 24/209 (bold dashed line),
the general shape of the observed H$\alpha$
profiles, including the larger
rise between 200-400\arcsec~ corresponding to our spectroscopic
positions B and C (see Fig.~1), is matched but with a 35\%
overprediction at around 230\arcsec.  In the lower panel, model
24/209 provides a nice fit to the observations except for
the predicted low ratios beyond 400{\arcsec}.  As noted above, the
heightened emission between 200-400{\arcsec} was reproduced by
raising the
total gas density within this region. Doing this, however,
caused the location of the
ionization front to move inward, causing a drop in emission
in both panels beyond 400\arcsec~ for model 24/209.  Attempts to
move the front out resulted in unacceptibly
smaller peaks in both panels in the 200\arcsec-400\arcsec~ region.
Finally, the X symbols in Fig.~6 correspond to values
derived from our spectral data in Table~2,
where positions A, B, and C are in order of
increasing distance from the central star.  These data are
reasonably consistent with the model as well as the imagery,
with the exception of the [N~II]/H$\alpha$ value for position C.

Fig. 7 provides additional support for our models but also
indicates some trouble spots.  The solid and dashed
lines show model predictions
for He~II $\lambda$4686, [O~II] $\lambda$3727, and [O~III]
$\lambda$5007, all relative to H$\beta$ and as a function of
distance from the central star in arcseconds.  Again, the X
symbols represent our spectral observations, which should be
compared with model 24/209.  We have also added
spectral results from O98 in order to obtain some constraints on
the inner region of the nebula.  His observations are shown with
horizontal lines bounded by vertical bars, where the line
centers correspond to positions of his slit centers and the line
lengths indicate the spatial extend of his
spectral extractions along the slit.  His slit position angle was
356$^{\circ}$, and thus his observations should be compared with
model 135/315.

We see in the top panel that He~II is observed to be strong close to
the central star but falls off to near zero at an angular distance of
$\sim$250\arcsec.  This is nicely reproduced by the models, where the
strong He~II has been produced by reducing the LOS thickness of the
nebula, which increases the relative amount of He$^{+2}$ to H$^+$ along
the LOS, and thus the ratio of associated line strengths goes up.  In
fact, we take the very strong $\lambda$4686/H$\beta$ as an indication
that the nebula is flat and not spherical for exactly this reason.  In
the middle and bottom panels, we see that qualitatively our models
reproduce the [O~II] and [O~III] observations: a low level of [O~II]
beginning near the central star with higher levels farther out, and the
reverse behavior occurring for [O~III].  Quantitatively, however, while
[O~II]/H$\beta$ and [O~III]/H$\beta$ are well matched by the models in
the inner and outer nebular regions, respectively, the model fails to
reproduce [O~II]/H$\beta$ beyond 200{\arcsec} and [O~III]/H$\beta$
inside of 200{\arcsec} by as much as a factor of two in each case.  The
problem could be resolved if a generally lower level of oxygen
ionization were present across the nebula, but trials with lower
stellar temperatures and higher gas densities caused unacceptable
changes in H$\alpha$ and [N~II] profiles in Fig.~6.  We conclude that
there are details of the nebular morphology and/or density structure
which are unaccounted for in our calculations, but testing more
sophisticated regimes is beyond the scope of this paper.

The ionization structures of He and O, electron temperature and density
behavior for the photoionization models associated with models 135/315
and 24/209, along with disk thickness profiles are shown in the five
panels of Fig.~8.  The top two panels track fractional ionization for
relevant ions of He and O across each model nebula, and since our final
LOS models are highly flattened, these panels closely represent the
behavior in ionization across the nebula in the plane of the sky.
Notice that in the vicinity of the central star the gas is dominated by
He$^{+2}$ and O$^{+3}$, consistent with O'Dell's empirical assessment.
Our models also support his finding of electron temperatures in excess
of 20,000~K in this same region, as shown in the third panel.  O'Dell
suggests that these high temperatures are generated by photoelectric
heating by dust, a point supported by our photoionization models which
include photoelectric effects that contribute close to 96\% of the
heating in the He$^{+2}$ region.  Note also that our model [O~III]
temperatures near positions A, B, and C are quite consistent with those
derived in the next subsection and shown here with X symbols.  The
ionization structure of the 24/209 model is radially compressed
relative to the 135/315 model, the result of the density enhancement
centered around 250{\arcsec} in the former model and appearing as the broad bump in the
fourth panel.  This density structure produces a more optically thick
nebula beginning in this region, consuming the ionizing photons within
smaller volumes.  Finally, the bottom panel in Fig.~8 shows values of
disk thickness (Z) relative to diameter (D) and may be thought of as an edge-on profile
of the Helix Nebula.  The bump between 200 and 300{\arcsec} of course
corresponds to the brightened ring region of the Helix.

In conclusion, our disk model calculations for the Helix Nebula,
inspired by O98 and produced by combining photoionization and LOS
models, is consistent with much available imagery and
spectrophotometric data, in particular those data for H$\alpha$ surface
brightness, and the profiles of [N~II] and He~II relative to hydrogen
emission.  Also, ionization structure, electron density and temperature
structure, and the edge-on profile all fit with observations, as do the
inferred stellar properties.  On the other hand, while the models do a
good job of reproducing the qualitative behavior for [O~II] and
[O~III], they overproduce emission in nebular regions where these lines
respectively dominate.  We emphasize that our modeling efforts here are only intended to test the viability of O'Dell's model of the Helix, and our model is not necessarily unique. Such a claim would require a more extensive exploration of parameter space, an exercise which is beyond the scope of this paper. However, if we assume the model's validity for the time being,
the model suggests that the Helix is generally
a filled disk with highly ionized material in the inner section which
often appears in images as a void.  The brightened ring results from a
thickening of the disk along the line-of-sight, while the regions
corresponding to our slit positions B and C are locations of density
enhancements and are thus additionally brightened.  We shall return to
a consideration of our model results in {\S}4.

\subsection{Empirical Abundances}

Abundances of He, C, N, O, Ne, S, and Ar at positions A, B, and C were derived
by combining our spectra with
a hybrid abundance method developed and employed in Henry, Kwitter, \& Howard 
(1996), and Kwitter \& Henry (1996; 1998).
The heart of the method is the use of a
photoionization model to improve results from a five-level atom
routine.  Briefly, we use a merged set of UV and optical line strengths and
derive an initial set of nebular abundance ratios.  We next
construct a nebular model which is tightly constrained by several important
observed diagnostics, and then calculate a second set of abundances
based upon the output linestrengths of the model.  The ratio of the actual
model input abundances to the abundances inferred from model output provides a
correction factor which is then applied to the original set of inferred
abundances to determine
the final set.

The above procedure can be expressed analytically for the abundance of one element X as
follows:
\begin{equation}
X = \left\{{\sum^{obs}}{{I_{\lambda}}\over{\epsilon_{\lambda}(T_e,N_e)}}\right\}
\cdot icf(X) \cdot \xi(X).
\end{equation}
$I_{\lambda}$ is the measured intensity of a spectral feature
produced by an ion of element $X$, $\epsilon_{\lambda} (T_e,N_e)$ is the
energy production rate per ion
of the spectral feature $\lambda$, icf(X) is the ionization correction factor, i.e. the ratio of the total abundances of all ions of X to the abundance sum of observable ions, and $\xi$(X) is the model-based correction factor alluded to above.  We now describe our procedural steps in greater detail.  

Emission lines used for calculating the ion abundance ratios inside the
curly brackets are He~I $\lambda$5876, He~II $\lambda$4686, [O~II]
$\lambda$3727, [O~III] $\lambda$5007,  [N~II] $\lambda$6584, C~III]
$\lambda$1909, [Ne~III] $\lambda$3869, [S~II]
$\lambda\lambda$6716,6731, [S~III] $\lambda\lambda$9069,9532, and
[Ar~III] $\lambda$7135.  Specific line strengths were taken from Table~2.
Values of $\epsilon$ are calculated using ABUN, which employs a
five-level atom routine, using atomic data the sources for which are listed in Table~3A.  This program also calculates electron
temperatures and densities.
Quotients for observable ions are summed together and multiplied by
the ionization correction factor (icf), where the ionization correction
factors are calculated according to the following prescriptions taken
from Kingsburgh \& Barlow (1994), with ion symbols, i.e. He$^{+2}$,
implying number abundances:
\begin{mathletters}
\begin{eqnarray}
icf(He) = {{He}\over{He^{+2}+He^+}} & = & 1.0, \\
icf(O)  =  {{O}\over{O^{+2}+O^+}} & = & {{He^+ + He^{+2}}\over{He^+}},\\
icf(C)  = {{C}\over{C^{+2}}} & = & {{O^+ + O^{+2}}\over{O^{+2}}} \cdot {{He^+ +
He^{+2}}\over{He^+}},\\
icf(N)  = {{N}\over{N^+}} & = & {{O^+ + O^{+2}}\over{O^+}} \cdot {{He^+ + 
He^{+2}}\over{He^+}},\\
icf(Ne)  = {{Ne}\over{Ne^{+2}}} & = & {{O^+ + O^{+2}}\over{O^{+2}}} \cdot {{He^+ 
+
He^{+2}}\over{He^+}},\\
icf(S)  = {{S}\over{S^{+2}+S^+}} & = & \left[1-\left(1-{{O^+}\over{O^+ + 
O^{+2}}} \cdot
{{He^+}\over{He^+ +
He^{+2}}}\right)^3\right]^{-0.33},\\
icf(Ar)  = {{Ar}\over{Ar^{+2}}} & = & 1.87.
\end{eqnarray}
\end{mathletters}

Table~3B lists the derived relative ion abundances, uncertainties, and icf's by position. We note that abundances of high ionization species such as He$^{+2}$, C$^{+2}$, O$^{+2}$, Ne$^{+2}$, and Ar$^{+2}$ were calculated using the [O~III] temperature, while abundances of He$^+$, O$^+$, N$^+$, S$^{+2}$, and S$^+$ were determined using the [N~II] temperature. As noted below, our [O~II] temperatures appear unreliable due to the uncertainty of the [O~II] $\lambda$7325 measurements, and thus we consistently used [N~II] temperatures for the lower ionization species.  
The uncertainties were determined rigorously by 
propagating assumed line strength uncertainties (see {\S}2.2.3) through the individual ion abundance calculations, i.e. the factors in curly brackets in Eq.~1, adding all differential contributions in quadrature.  Thus, effects of electron temperature uncertainties on the abundances are accurately accounted for.  We note, however, that contributions from uncertainty in atomic data were not included in the error analysis.  We also caution that the actual uncertainty in the C$^{+2}$ abundance at position~C is greater than indicated, due to the absence of He~II $\lambda$1640 at position~C, which prohibited the determination of a UV-optical merging factor at that location.
It is interesting to note that as distance from the central star increases in going from position~A to position~C, the abundances of higher ionized species, i.e. O$^{+2}$, Ne$^{+2}$, and S$^{+2}$, are seen to systematically decrease, while those of O$^+$, N$^+$, and S$^+$ appear to increase, as is expected because of the decreased density of ionizing photons in more distant regions of the nebula. 

We thus derive a preliminary abundance for an element by summing the observed ion abundances for that element in Table~3B and then multiplying the sum by the appropriate ionization correction factor.

The final step in our procedure is to multiply the preliminary abundances by a correction factor
$\xi(X)$, defined as:  \begin{equation} 
{{true~model~abundance}\over{apparent~model~abundance}}.
\end{equation} The correction factor $\xi$ is a gauge of the accuracy
of the use of the ionization correction factor method for determining
abundances and helps correct abundances determined in the traditional method for effects such as charge exchange. The correction factor is determined by using CLOUDY 90.03a (Ferland 1990) to
calculate a photoionization model which best reproduces a wide range of
observed diagnostic line ratios.  The line strength output by the model
is then used as input to ABUN to derive a set of abundances, i.e. the
{\it apparent} abundances in eq.~3.  These are compared with the input
or {\it true} abundances for the model to yield $\xi$.  In particular,
the photoionization models for this segment of the abundance
determinations were calculated for each slit position in order to
reproduce as closely as possible the physical conditions observed along
the line-of-sight.  Our models were constrained by a set of 10
important diagnostic ratios constructed directly from observed line
strengths.  These 10 ratios are known to describe the physical
conditions of a nebula quite well.  Our goal for each of the three
positions was to match each observed ratio to within 0.10-0.15~dex,
consistent with observational uncertainties\footnote{A possible
alternative to this method would be to use our model results in the
last section to calculate ionization correction factors along
lines-of-sight coincident with our three observed positions.  However,
while the models in {\S}3.2 successfully reproduced most of the
observations, they did not match the [O~II] $\lambda$3727 and [O~III]
$\lambda$5007 emission satisfactorily throughout the nebula.
Therefore, we decided to continue using the routine we have used in the
three previous papers which are part of this series on PN
abundances.}.  We assumed that the central stars were blackbodies and
that the nebula had a uniform density with a filling factor of
unity.{\footnote{While most PNe are known to have filling factors
significantly less than unity, the only measurable quantity affected by
the filling factor is the nebular luminosity, a parameter we are not
using to constrain our models.  Thus, using the same filling factor for
all of the nebulae in no significant way influences our abundance
results.}}  The inner nebular radius was taken to be 0.032~pc for all
models, but the outer radius was treated as a free parameter.  In
several cases the best matches to the observed line strengths were
produced by truncating the model inside the Str{\"o}mgren radius, i.e.
the model nebula was matter-bounded.  Other free parameters included
the stellar luminosity, nebular electron density, and nebular
abundances of helium, oxygen, nitrogen, carbon, neon, and sulfur.

Table~4A summarizes our model results; for each position we list
logarithmically the observed and model-predicted values
for 10 important diagnostic line ratios in the upper section of the
table.  The first ratio is sensitive to gas-phase metallicity and
electron temperature, the second and third to the level of nebular
excitation, the fourth and fifth to electron temperature and density,
respectively, and the last five to abundance ratios in the order He/H,
N/O, S/O, C/O, and Ne/O.  The lower section of the table provides
important model input parameters: stellar effective temperature
(T$_{eff}$), the log of stellar luminosity log($L$),
electron density (N$_e$), and the inner and outer nebular
radii (R$_o$ and R; values of R which are less then the Str{\"o}mgren
distance, i.e. matter bounded models, are indicated with a
footnote). These are followed by six input abundance ratios.
(N.B. We emphasize that these
abundance ratios are {\it not} our final abundances for
each object, but are the abundances necessary to produce the best
model.)

There are several important points about the model results that require
discussion.  First, we note that these models represent the best-fit
solutions to a specific line-of-sight position within a nebula; they
are {\it not} models of whole planetary nebulae.  Model parameters were
varied to match quantities measured along a single line-of-sight where
local conditions may have large effects on the line strengths.  Thus,
none of the values for these parameters is necessarily the same as the
actual one.  Rather, whole nebula properties are treated above in
{\S}3.2 using our narrow-band filter images.

We draw the reader's attention to the satisfactory agreement between
observation and theory in all line ratios except the one associated
with He~II/He~I for position~C, where the model grossly overpredicts
the line strength.  This ratio is particularly sensitive to nebular
excitation, as is the [O~II]/[O~III] ratio above it.  Curiously, we
were able to obtain a good match for the latter, as well as all of the
other ratios at this position, suggesting that perhaps our measured
line ratio is spurious.  We note that the strength of the He~II/He~I
ratio at the other two positions is much higher.

The correction factors $\xi$ derived using our models and eq.~3
are given in Table~4B.  Since all but one of the values is less than unity, we
conclude that generally the standard abundance method of using a 5-level atom
calculation along with an icf results in values which are systematically too
large, although
never by more than a factor of two in the worst cases.

The one instance in which we deviated from the above procedure was in the case of the sulfur abundance for position~A where [S~III] was not measured. Here we used the relation between S$^+$ and S$^{+2}$ derived by Kingsburgh \& Barlow (1994), i.e. $S^{+2}/S^+ = 4.677 + (O^{+2}/O^+)^{0.433}$, to obtain an estimate of the $S^{+2}$ abundance. We then determined a value for $\xi$ from our model of position~A using this same relation, and obtained a value for S/H. We checked the consistency of this method by applying it to positions B and C and found abundances determined in this way to be very consistent with those calculated using the method expressed in eq.~1.

Our final abundance results for the three positions in the Helix are
given in Table~5, where we include our derived values for [O~III], [N~II], and [O~II] electron temperatures\footnote{The [O~II] temperatures are significantly different among the three positions and are inconsistent with [N~II] temperatures as well as our model predictions. Presumably, this is due to the large uncertainties in our measurements of the $\lambda$7325 quartet feature. As noted earlier, we have opted to use [N~II] rather than [O~II] temperatures to calculate abundances of lower ionization species.} and [S~II] electron densities.  We
also show average values for our three positions (with standard deviations), as well as
results for the Helix from Peimbert, Luridiana, \& Torres-Peimbert
(1995), Hawley (1978), and O'Dell (1998), averages for a large sample of PNe by
Kingsburgh \& Barlow (1994), abundances in Orion (Esteban et al. 1998),
and solar abundances from Grevesse, Noels, \& Sauval (1996).  Uncertainties in all quantities have been determined by estimating contributions from the ionization correction and $\xi$ factors and adding them together in quadrature along with uncertainties in ion abundances given in Table~3B. 

We find the Helix Nebula to have a metallicity somewhat less than
solar, as implied by O/H, in agreement with Hawley but not with
Peimbert et al.  Our average O/H is similar to the value found by
Kingsburgh \& Barlow (1994) for a large sample of southern PNe as well
as for the O/H level found by Esteban et al. (1998) for the Orion
Nebula.  Also, O/H in the Dumbbell Nebula (NGC~6853), another PN of
close proximity, i.e. 240~pc (Harris et al. 1997), is found by
Perinotto (1991) to be 4.1$\times 10^{-4}$, close to our value for the
Helix.  We conclude, then, that O/H in the Helix is largely
representative of the local ISM level.  With a mass of 6.5M$_{\sun}$ (see {\S}4), the Helix progenitor formed less than
100 million years ago, when the ISM metallicity
would have been similar to its current level. Therefore, the
agreement between the Helix and local ISM metallicities is not
unexpected.

Our carbon abundances in Table~5 are the first of their kind published for the 
Helix.  The uncertainties for carbon are relatively large primarily because of the temperature sensitivity of the excitation rate of [C~III] $\lambda$1909.  Unfortunately, the recombination line C~II $\lambda$4254 was not observed in the Helix, and thus we were unable to estimate a carbon abundance using a less temperature-sensitive method.  Note that the average C/O value quoted pertains to positions~A and B only, since no UV-merging factor could be determined at position~C.
Considering the uncertainties, the values we 
find for C/O appear consistent with the average PN level of 
Kingsburgh \& Barlow, as well as the values for Orion and the sun.   We are therefore unable 
to state conclusively that carbon is enriched.   

Certainly the evidence for nitrogen enrichment is much greater than for carbon 
enrichment in the Helix.  Our average level of 0.54 for N/O is well above both the solar 
and Orion levels.  In fact the Helix fits the definition of a Peimbert Type~I 
PN, with N/O$\ge$0.5 and He/H$\ge$0.12.

Our results for Ne/O and Ar/O show consistency with other PNe, i.e. the KB 
sample, Orion, and marginally with the sun.  According to the stellar yields of 
Woosley \& Weaver (1995) $^{16}$O, $^{20}$Ne and $^{36}$Ar are synthesized by 
stars within the same general mass range, i.e. stars with birth masses greater 
than 15M$_{\sun}$, and thus Ne/O and Ar/O should be roughly the same everywhere in space and time. Since PN progenitors are not expected to alter their 
inherited values of Ne and Ar (and probably O, unless ON cycling is 
significant), it is not too surprising that the two ratios should be the same 
everywhere.  Our results seem consistent with that idea.

The main challenge in our abundance results is attempting to understand
the origin of the very low S/O value.  We find values roughly an order
of magnitude below the KB average for PNe, the sun, and Orion.  It
should be noted that a few objects in the KB sample have levels of S/O
similar to ours, and it is clear from the data that these low S/O
levels in both the Helix and the KB sample are associated with
relatively weak [S~II] emission observed in all of these object.
Because the $\xi$ corrections are within 25\% of unity, we are
reasonably confident of our abundance method for deriving S in the
Helix.  Yet, since S and O are both produced in massive stars, the S/O
level in a planetary nebula must represent the value which was present
in the interstellar medium at the time that the PN progenitor formed. A
depressed value of S/O in the ISM could have been present if, shortly
before the Helix progenitor formed nearby, an evolved massive star had
ejected material which was low in sulfur relative to oxygen. This
nucleosynthetic mix is possible if the boundary between stellar core
material which is to be ejected and the material which remains behind
in the remnant, i.e. the mass cut, is more distant from the center of
the massive star than normal (Nomoto et al. 1997). However, such a
reduced sulfur level should be accompanied by a similar reduction in
argon, yet that is not inferred from our observations.  Another
possible explanation is that significant amounts of sulfur reside in
ionization states above S$^{+2}$, i.e. states which we could not
observe directly or account for because of physical conditions
unrepresented in our models.

In summary, the Helix Nebula appears to be a Type~I PN with He and N enrichment and a metallicity consistent with that of the
solar neighborhood as measured by other nebulae, although it is lower
in metallicity than the sun.  The level of carbon relative to oxygen in the Helix seems
to be typical of what is seen in local PNe and H~II regions, although C/O
exceeds the solar value.

\section{Discussion}

The relative success of our whole-nebula models along with our abundance 
measurements of the Helix Nebula provide us with an opportunity to assemble a 
composite picture of the past and present of this nearby planetary nebula.  
These models suggest a central star of T$_{eff}$=120,000K and a 
luminosity of 100L$_{\sun}$, implying a radius of 0.02R$_{\sun}$.  The values 
for T$_{eff}$ and L are very similar to published values quoted in the 
introduction, while the radius is consistent with the central star 
classification as a DAO white dwarf (Napiwotzki \& Sch{\"o}nberner 1995). 
G{\'o}rny et al. (1997) estimate the mass of the central star to be 
0.93M$_{\sun}$, which, when combined with the initial-final mass relation of (Weidemann 1987), suggests a progenitor mass of 6.5M$_{\sun}$, corresponding to 
a B5 main sequence star.  At its current distance the Helix progenitor 
would have appeared in our sky with a visual magnitude of 5.5, barely 
discernible with the naked eye.  Its main sequence lifetime is 
estimated to have been 56~million years, using the information for stellar 
lifetimes in Schaller et al. (1992).  The dynamical age of the nebula is given 
by G{\'o}rny et al. as 22,200 years, based upon Harris' 
(1997) distance and a linear diameter of roughly 1~pc, implying an old planetary nebula and explaining its relatively low H$\alpha$ surface 
brightness.

The results of our whole-nebula models support the disk morphology of
the Helix proposed by O98. In this model the Helix Nebula is imagined
to be a disk generally 0.33~pc in thickness which is inclined about
30{\arcdeg} with respect to the plane of the sky around an axis with a
position angle of 22{\arcdeg}.  The disk model and the visible presence
of a ring of enhanced emission is consistent with the idea of an
isotropic wind interacting with and being directed by a remnant red
giant envelope which tends to be concentrated in an equatorial plane
(see Balick 1987).  The inner region of the disk out to a radial
distance of 0.2~pc contains hot gas heated mostly by photoelectric
heating through dust in which the temperature exceeds 20,000K.  This
region is also dominated by highly ionized material such as He$^+$ and
O$^{+3}$, making it appear empty in the light of lower ionization
species such as [N~II], [O~II], [O~III], or [S~II].  The total density
of this region, as well as for the most of the nebula, is estimated to
be around 60~cm$^{-3}$.

Beyond the inner region between 0.2 and 0.3~pc radially is the bright
ring, the assumed interaction site of a fast stellar wind with the
previously-ejected red giant envelope (Balick 1987).  Here our models
indicate that the enhanced emission is explained by a near doubling of
the disk thickness, producing a greater column density of material; an
increase in particle volume density in this region is not suggested.  However, in the
bright regions where our spectroscopic observations were made, the
enhanced brightness above the level of the ring is consistent with a
density of roughly 120~cm$^{-3}$, i.e. double the surrounding value.
The electron temperature in the ring is estimated to be around
9,000-10,000K, where the heating is no longer controlled by
photoelectric processes but by photoionization. The ring region is
dominated by He$^{+}$, O$^{+2}$, O$^{+}$, and N$^{+}$.

Outside of the ring, the ionization level continues to drop, with the
ionization front located at around 0.45~pc.  Beyond this point, the
gas is dominated by neutral atomic species.  If we assume a disk radius
of 0.6~pc, a total gas density of 60~cm$^{-3}$, a filling factor of 0.55 (inferred from models in {\S}3.2), and a disk thickness of
0.33~pc, we estimate a nebular mass of 0.30M$_{\sun}$, which
constitutes a lower limit, since presumably the nebular gas extends
beyond the 0.6~pc distance.  Assuming a progenitor birth mass of
6.5M$_{\sun}$ and a stellar remnant mass of 0.93M$_{\sun}$, this implies that
roughly 5.3M$_{\sun}$ of stellar material is currently dispersed or
otherwise unobservable.

Finally, to speculate a bit, during its lifetime, the progenitor of the
Helix, given its mass, experienced three dredge-up phases (Iben 1995).
The first two took place while the star was on the red giant branch and
early on the asymptotic giant branch, respectively, and deposited
$^4$He and $^{14}$N in the envelope, while the third phase, which
occurs late in the AGB phase, is characterized by thermal pulses and
would have dredged up $^{12}$C into the atmosphere.  To the extent that
hot bottom burning occurred, some of the $^{12}$C would have been
converted to $^{14}$N during this phase, thus reducing the amount of
the former in the envelope.  According to stellar evolution models by
van~den~Hoek \& Groenewegen (1997), hot bottom burning occurs in AGB
stars of masses in excess of 5M$_{\sun}$.  Therefore, the evidence
presented here for minimal carbon enrichment and modest nitrogen
enrichment is consistent with the hot-bottom burning prediction,
assuming that the Helix progenitor's mass was 6.5M$_{\sun}$.

\section{Summary}

We have undertaken a comprehensive study of NGC~7293, the Helix Nebula,
a study which had two goals: (1) to learn about the morphology of the
ionized nebula; and (2) to determine accurate chemical abundances in
the nebula.  In the process, we have obtained, and present here, new
narrow-band filter imagery as well as UV and optical spectrophotometry
extending from around 1200{\AA} to 9600{\AA}.

The first part of the study involved the use of photoionization models
coupled with line-of-sight integrations of emissivities from the former
in an attempt to reproduce the observed profiles of H$\alpha$ surface
brightness and [N~II]/H$\alpha$ ratios.  We were able to match our own
observations in addition to those of O'Dell (1998) reasonably well. In
addition, our new optical spectrophotometry permitted us to estimate
the abundances of He, C, N, O, Ne, S, and Ar at three bright points in
the nebula.  To our knowledge, this is the first time such measurements
have been made in the Helix for C and S.

We have arrived at the following results:

\begin{enumerate}

\item Our H$\alpha$ imagery and photoionization models of the Helix Nebula support 
the disk model proposed by O'Dell 
(1998), which includes a hot, highly ionized inner region heated 
largely by photoelectric processes, as well as a cooler, lower-ionization ring 
region.

\item Our models also support a central star effective temperature of 
120,000K and a luminosity of 100L$_{\sun}$.

\item The visible disk is roughly 1~pc in diameter and 0.33~pc thick, and is 
tipped with respect to the sky plane by 30{\arcdeg} around an axis with position 
angle of 22{\arcdeg}, where these angles agree well with those found by O'Dell 
and Meaburn et al. (1998).  A lower limit to the mass of the visible nebula is 
0.55M$_{\sun}$.

\item Abundance measurements suggest that the nebula is enhanced in He
and N and we confirm its earlier classification as a Type~I PN.  Within
the uncertainties, carbon does not appear to be enriched relative to
oxygen, suggesting that hot-bottom burning occurred during the AGB
phase of the progenitor's evolution.   The Helix also appears to have
an anomalously low sulfur abundance, which is seen in a few other PNe,
but is nevertheless difficult to understand currently in light of
current nucleosynthesis scenarios.

\end{enumerate}

\acknowledgments

We would like to acknowledge support from the staff at KPNO and Palomar
Observatory, as well as at the IUE.  R.B.C.H. and K.B.K thank their
respective institutions for travel support to KPNO.  R.J.D. and
R.B.C.H. are grateful to the University of Oklahoma and Rice
University, respectively, for hospitality during visits.  R.J.D. wishes
to thank J. Hester for use of the WF PFUEI camera on the Palomar 1.5m
to obtain the imagery presented here and to NASA/IUE grant NAG5-262 for
supporting the original IUE observations. Finally, we would like to thank Bob O'Dell and Ken Nomoto for useful conversations, and an anonymous referee who offered numerous suggestions for improving the paper.

\appendix

\section{Line-Of-Sight Integration}

The observed nebular surface profiles shown in Fig.~6 were modeled in
two steps.  First, a photoionzation model (CLOUDY) using spherical
symmetry was calculated by assuming a set of input nebular and stellar
properties.  Relevant output volume emissivities as a function of
radial distance were then used as input to a separate program written
by one of us (R.B.C.H.) which computed line-of-sight intensities at
regular points in the nebula.

The program began by computing the pathlength $S$
through the nebula of angular diameter $\Theta$ along a given
line of sight located at an angular distance $\alpha$ from the central star\footnote{Angular measurements are easily converted to parsecs or
kilometers once the distance is known.  In our case we assumed a
distance to the Helix of 213~pc (Harris 1997).}.
Thus:
\begin{equation}
S = 2 \times \left\{\Theta^2-\alpha^2\right\}^{1/2}
\end{equation}
The pathlength $S$ was then divided into 1000 equal
segments, and beginning at the proximal surface of the nebula
the program stepped along $S$ through the gas, and
at each point $s$ added the emissivity contribution appropriate for the radial 
distance $r$:
\begin{equation}
r = \left\{\alpha^2 + \left({{S}\over{2}}-s\right)^2\right\}^{1/2}
\end{equation}
where $s$ is the distance within the nebula along the
line-of-sight. This calculation was repeated at 10\arcsec~ intervals beginning at the central star and extending outward to the edge of the nebula.  Adjustments were made for the empty nature of the region inside the inner nebular edge.  However, because the Helix appears to be tipped by only about 30\arcdeg~ with respect to the sky plane, the small increase in line-of-sight pathlength which results was ignored.

\begin{deluxetable}{lllll}
\tablecolumns{5}
\tablewidth{0pc}
\tablenum{1A}
\tablecaption{Summary of Imagery Observations}
\tablehead{
\colhead{Date (UT)} &
\colhead{Instrument} &
\colhead{Filter} &
\colhead{$\lambda_o$/$\Delta\lambda$\tablenotemark{1}} &
\colhead{Exp(sec)}
}
\startdata
2 Dec 1988 & 1.5m WF PFUEI & Halpha & 6563/15 &    600 \\
&&        RC1  &   6450/104 &   120\\
&&        [NII] &  6584/14  &   900\\
&&        [SII]  & 6730/36  &  1200\\
&&        IRC1   &  7230/100  &   240
\enddata
\tablenotetext{1}{$\lambda_o$ = central wavelength,
$\Delta\lambda$ = FWHM, both in \AA.}
\end{deluxetable}

\begin{deluxetable}{lcccc}
\tablecolumns{5}
\tablewidth{0pc}
\tablenum{1B}
\tablecaption{Summary of UV Spectrophotometric Observations}
\tablehead{
\colhead{Position} &
\colhead{Offsets\tablenotemark{1}} &
\colhead{Date (UT)} &
\colhead{Grating} &
\colhead{Exp(sec)}
}
\startdata
A & 97E, 171N & 3 Oct 1991 & SWP42615 & 31200\\
&& 3 Oct 1991 & LWP21398 & 28200\\ 
B & 92E, 212N & 25 Nov 1990 & LWP19278 & 9000\\
&& 25 Nov 1990 & SWP40195 & 3600\\
&& 25 Nov 1990 & SWP40196 & 14400\\
C & 151W, 269S & 4 Oct 1991 & SWP42621 & 25200\\
&& 4 Oct 1991 & LWP21401 & 22800
\enddata
\tablenotetext{1}{Given in arcseconds in the sky plane (uncorrected
for nebular inclination) with respect to the central star.  The offsets are for 
the slit center.  Slit position angle for both A and C was 309\arcdeg~ and 322\arcdeg~ for B.}
\end{deluxetable}

\begin{deluxetable}{lcccc}
\tablecolumns{5}
\tablewidth{0pc}
\tablenum{1C}
\tablecaption{Summary of Optical Spectrophotometric Observations}
\tablehead{
\colhead{Position} &
\colhead{Offset\tablenotemark{1}} &
\colhead{Date(UT)} &
\colhead{Grating\tablenotemark{2}} &
\colhead{Exp(sec)}
}
\startdata
A & 97E, 171N & 7 Dec 1996 & Red & 1800\\
&& 8 Dec 1996 & Blue & 1800\\
B & 92E, 212N & 9 Dec 1996 & Red & 900\\
&&8 Dec 1996 & Blue & 1200\\
C & 151W, 269S & 9 Dec 1996 & Red & 1800\\
&&8 Dec 1996 & Blue & 1200
\enddata
\tablenotetext{1}{Given in arcseconds in the sky plane
(uncorrected for nebular inclination) with respect to the central star. The 
offsets are for the slit center. The slit position angle in each case was 
90{\arcdeg}.}
\tablenotetext{2}{Red: KPNO grating \#58, blazed at 8000\AA; Blue: KPNO
grating \#240, blazed at 5500\AA.}
\end{deluxetable}

\begin{deluxetable}{lrrrrrrr}
\small
\vspace{-2in}
\tablecolumns{8}
\tablewidth{0pc}
\tablenum{2A}
\tablecaption{UV \& Optical Line Strengths}
\tablehead{
\colhead{} & \colhead{} &
\multicolumn{2}{c}{A} &
\multicolumn{2}{c}{B} &
\multicolumn{2}{c}{C} \nl
\cline{3-4} \cline{5-6} \cline{7-8} \nl
\colhead{Line} &
\colhead{f$_{\lambda}$} &
\colhead{F$_{\lambda}$} &
\colhead{I$_{\lambda}$} &
\colhead{F$_{\lambda}$} &
\colhead{I$_{\lambda}$} &
\colhead{F$_{\lambda}$} &
\colhead{I$_{\lambda}$}
}
\startdata
C II $\lambda$1336 &1.41&\nodata&\nodata&7.8 &10 &\nodata&\nodata\nl
N IV] $\lambda$1485 &1.23&\nodata&\nodata&11 &14 &\nodata&\nodata\nl
C IV $\lambda$1549 &1.18&\nodata&\nodata&\nodata&\nodata&56 &56 \nl
He II $\lambda$1640 &1.14&30 &57 &17 &21&\nodata&\nodata\nl
N III] $\lambda$1750 &1.12&&&9.5 &11 &38 &38 \nl
C III] $\lambda$1909 &1.23&37 &70 &50&63 &95 &95 \nl
\[[O III] + C II] $\lambda$2325 &1.35&66 &123 &25 &33 &65 &65 \nl
\[[O II] $\lambda$ 3727 &0.29&188 &188 &475 &518 &718 &718 \nl
He II + H10 $\lambda$3797 &0.27&\nodata&\nodata&\nodata&\nodata&5.6 &5.6 \nl
He II + H9 $\lambda$3835&0.25&\nodata&\nodata&5.9 &6.4 &7.1 &7.1 \nl
\[[Ne III] $\lambda$3869 &0.25&137 &137 &101 &109 &97 &97 \nl
He I + H8 $\lambda$3889 &0.25&21 &21 &24 &26 &25 &25 \nl
H$\epsilon$ + \[[Ne III] $\lambda$3968 &0.23&71 &71 &78 &84 &80 &80 \nl
He II $\lambda$4026 &0.21&\nodata&\nodata&1.1 &1.2 &1.6 &1.6 \nl
\[[S II] $\lambda$4072 &0.20&\nodata&\nodata&1.5 &1.6 &1.7 &1.7 \nl
He II + H$\delta$ $\lambda$4101 &0.19&23 &23 &22 &23 &23 &23 \nl
H$\gamma$ $\lambda$4340 &0.13&46 &46 &46 &47 &44 &44 \nl
\[[O III] $\lambda$4363 &0.12&3.5 &3.5 &2.6 &2.7 &1.7 &1.7 \nl
He I $\lambda$4471 &0.09&5.1 &5.1 &5.5 &5.6  &5.6 &5.6 \nl
He II $\lambda$4686 &0.04&8.9 &8.9&3.3 &3.3 &0.46 &0.46 \nl
H$\beta$ $\lambda$4861 &0.00&100 &100 &100 &100 &100 &100  \nl
He I $\lambda$4922 &-0.02&2.2 &2.2 &1.9 &1.9 &1.7 &1.7 \nl
\[[O III] $\lambda$4959 &-0.03&209 &209 &168 &167 &81 &81  \nl
\[[O III] $\lambda$5007 &-0.04&714 &714 &513 &507 &309 &309 \nl
\[[N II] $\lambda$5755 &-0.21&3.2 &3.2 &6.3 &6.0 &11 &11 \nl
He I $\lambda$5876 &-0.23&15 &15 &19 &17 &17 &17 \nl
\[[N II] $\lambda$6548 &-0.36&81 &81 &167 &150 &280 &280 \nl
H$\alpha$ $\lambda$6563 &-0.36&272 &272 &318 &286 &284 &284 \nl
\[[N II] $\lambda$6584 &-0.36&250 &250 &517 &465 &842 &842 \nl
He I $\lambda$6678 &-0.38&4.2 &4.2 &5.5 &4.9 &4.9 &4.9 \nl
\[[S II] $\lambda$6716 &-0.39&3.6 &3.6 &10 &9.2 &22 &22 \nl
\[[S II] $\lambda$6731 &-0.39&2.6 &2.6 &7.4 &6.6 &16 &16 \nl
He I $\lambda$7065 &-0.44&3.1 &3.1 &3.4 &3.0 &3.5 &3.5 \nl
\[[Ar III] $\lambda$7135 &-0.45&23 &23 &31 &27 &25 &25 \nl
\[[O II] $\lambda$7325 &-0.48&6.4: &6.4: &7.1: &6.2: &11 &11 \nl
\[[Ar III] $\lambda$7751 &-0.54&6.3 &6.3 &6.6 &5.7 &7.1 &7.1 \nl
\[[S III] $\lambda$9069 &-0.67&\nodata&\nodata&0.99: &0.81: &4.2: &4.2: \nl
P8 $\lambda$9228 &-0.68&4.1 &4.1 &\nodata&\nodata&4.0 &4.0 \nl
\[[S III] $\lambda$9532 &-0.70&\nodata&\nodata&28: &23: &16: &16: \nl
log F$_{H\beta}$\tablenotemark{a} &&-12.90&&-12.79&&-12.84&\nl
c &&&0.00&&0.13&&0.00\nl
merging factor\tablenotemark{b} &&&1.88&&0.87&&1.00\nl
\enddata
\tablenotetext{a}{Ergs/cm$^2$/s in our extracted spectra}
\tablenotetext{b}{Factor by which dereddened UV line strengths are multiplied
in order to merge them with optical data (see text).}
\end{deluxetable}

\begin{deluxetable}{lccccccc}
\tablecolumns{8}
\tablewidth{0pc}
\tablenum{2B}
\tablecaption{Line Ratios}
\tablehead{
\colhead{} & \colhead{} & \multicolumn{6}{c}{Observed}\\ \cline{3-8} \nl
\colhead{Ratio} & \colhead{Theory} & \colhead{A} & \colhead{$\Delta$\tablenotemark{a}} & \colhead{B} & \colhead{$\Delta$\tablenotemark{a}} & \colhead{C} &
\colhead{$\Delta$\tablenotemark{a}}
}
\startdata
[Ne~III] 3869/3968\tablenotemark{b} & 3.32 & 2.49 & 25.0 & 1.60 & 51.8 & 1.52 & 54.2 \nl
He~I 5876/4471 & 2.76 & 2.94 & 6.5 & 3.04 & 10.1 & 3.04 & 10.1 \nl
[O~III] 5007/4959 & 2.89 & 3.42 & 18.3 & 3.04 & 5.2 & 3.81 & 31.8 \nl
[N~II] 6584/6548 & 2.95 & 3.08 & 4.6 & 3.10 & 5.1 & 3.01 & 1.9 \nl
He~I 6678/4471 & 0.79 & 0.82 & 4.2 & 0.88 & 10.8 & 0.88 & 10.8 \nl
[Ar~III] 7135/7751 & 4.14 & 3.65 & 11.8 & 4.74 & 14.4 & 3.52 & 14.9 \nl
P8/H$\beta$ 9228/4861 & 0.037 & 0.041 & 10.8 & \nodata & \nodata & 0.04 & 8.1 \nl
[S~III] 9532/9069 & 2.48 & \nodata & \nodata & 28.4 & 1044 & 3.81 & 53.6 \nl
\enddata
\tablenotetext{a}{$\Delta$=${{\vert Observed-Theory \vert}\over{Theory}} \times 100.$}
\tablenotetext{b}{The [Ne~III] $\lambda$3968 line was corrected for the contribution from H$\epsilon$.}
\end{deluxetable}

\begin{deluxetable}{lcl}
\small
\vspace{-2in}
\tablecolumns{3}
\tablewidth{0pc}
\tablenum{3A}
\tablecaption{ABUN: Sources Of Atomic Data}
\tablehead{
Ion&Data Type\tablenotemark{a}&Reference
}
\startdata
H$^0$ & $\alpha_{eff}$($\lambda$4861) & Storey \& Hummer 1995 \nl
He$^0$ & $\alpha_{eff}$($\lambda$5876)\tablenotemark{b} & P{\'e}quignot et al.
1991 \nl
He$^+$ & $\alpha_{eff}$($\lambda$4686) & Storey \& Hummer 1995 \nl
O$^+$ & $\Omega$ & Mendoza 1983 (2-3,4-5); McLaughlin \& Bell 1993 (all other
transitions) \nl
 & A & Wiese, Fuhr, \& Deters 1996 \nl
O$^{+2}$ & $\Omega$ & Burke, Lennon, \& Seaton 1989 (4-5); Lennon \& Burke
1994 (all other transitions) \nl
 & A & Wiese, Fuhr, \& Deters 1996 \nl
N$^+$ &  $\Omega$ & Lennon \& Burke 1994 \nl
 & A & Wiese, Fuhr, \& Deters 1996 \nl
C$^{+2}$ & $\Omega$ & Berrington et al. 1985 \nl
 & A & Nussbaumer \& Storey 1978; Kwong et al. 1993 ($\lambda$1909 only) \nl
Ne $^{+2}$ & $\Omega$ & Butler \& Zeippen 1994 \nl
 & A & Baluja \& Zeippen 1988 \nl
S $^{+}$ & $\Omega$ & Ramsbottom, Bell, \& Stafford 1996 \nl
 & A & Mendoza 1983 \nl
S $^{+2}$ & $\Omega$ & Galav{\'i}s et al. 1995 \nl
 & A & Mendoza 1983 \nl
Ar $^{+2}$ & $\Omega$ & Galav{\'i}s et al. 1995 \nl
 & A & Mendoza \& Zeippen 1983 \nl
\enddata
\tablenotetext{a}{$\alpha_{eff}$=effective recombination coefficient;
$\Omega$=collision strength; A=transition rate.}
\tablenotetext{b}{Includes collisional effects given by Clegg (1987).}
\end{deluxetable}

\begin{deluxetable}{llll}
\tablecolumns{4}
\tablewidth{0pc}
\tablenum{3B}
\tablecaption{Ion Abundances \& Ionization Correction Factors}
\tablehead{
\colhead{Ion Ratio} &
\colhead{A} &
\colhead{B} &
\colhead{C}
}
\startdata
He$^+$/H$^+$ & 0.11($\pm$0.015) &0.13($\pm$0.016) & 0.13($\pm$0.016)\nl
He$^{+2}$/H$^+$($\times 10^{3}$) & 7.89($\pm$0.74)&2.97($\pm$0.27)&0.41($\pm$0.04) \nl
icf(He) & 1.00 &1.00 &1.00 \nl\nl
O$^+$/H$^+$($\times 10^{4}$) & 0.94($\pm$0.26) &2.38($\pm$0.87) &3.27($\pm$0.91) \nl
O$^{+2}$/H$^+$($\times 10^{4}$) & 3.62($\pm$0.86) &2.25($\pm$0.30) &1.38($\pm$0.20) \nl
icf(O) & 1.07 & 1.02&1.00 \nl\nl
C$^{+2}$/H$^+$($\times 10^{4}$) & 3.74($\pm$6.4) &2.52($\pm$2.2) & \nodata\nl
icf(C) & 1.35 & 2.10 & \nodata \nl\nl
N$^+$/H$^+$($\times 10^{4}$) & 0.62($\pm$0.06) &1.04($\pm$0.13) &1.84($\pm$0.20) \nl
icf(N) & 5.18 &1.99 &1.43 \nl\nl
Ne$^{+2}$/H$^+$($\times 10^{4}$) & 2.04($\pm$0.68) &1.43($\pm$0.24) &1.29($\pm$0.24) \nl
icf(Ne) & 1.35 &2.10 &3.37 \nl\nl
S$^+$/H$^+$($\times 10^{7}$) & 1.77($\pm$0.35) &4.03($\pm$1.1) &9.61($\pm$2.1) \nl
S$^{+2}$/H$^+$($\times 10^{6}$) & \nodata &1.45($\pm$0.57) &1.04($\pm$0.42) \nl
icf(S) & 1.28 &1.04 &1.01 \nl\nl
Ar$^{+2}$/H$^+$($\times 10^{6}$) & 2.90($\pm$0.47) &2.88($\pm$0.33) &2.66($\pm$0.32) \nl
icf(Ar) & 1.87 &1.87 &1.87
\enddata
\end{deluxetable}

\begin{deluxetable}{lrrrrrr}
\tablecolumns{7}
\tablewidth{0pc}
\tablenum{4A}
\tablecaption{Observations \& Models}
\tablehead{
\colhead{} &
\multicolumn{2}{c}{A} &
\multicolumn{2}{c}{B} &
\multicolumn{2}{c}{C} \nl
\colhead{} &
\colhead{Obs} &
\colhead{Model} &
\colhead{Obs} &
\colhead{Model} &
\colhead{Obs} &
\colhead{Model}
}
\startdata
log ($I_{[O II]}+I_{[O III]}$)/H$\beta$ &1.06&1.10&1.08&1.01&1.06&1.04\nl
log $I_{[O II]}/I_{[O III]}$ &-0.71&-0.71&-0.12&-0.13&0.24&0.24\nl
log I$_{He II/He I}$ &-0.22&-0.22&-0.72&-0.72&-1.57&-0.17\nl
log I$_{\lambda 4363}$/I$_{\lambda 5007}$ &-2.31&-2.32&-2.27&-2.24&-2.27&-
2.22\nl
log I$_{\lambda 6716}$/I$_{\lambda 6731}$ &0.14&0.14&0.15&0.15&0.14&0.15\nl
log I$_{He~I/H\beta}$&-0.83&-0.79&-0.76&-0.77&-0.77&-0.78\nl
log I$_{6584/3727}$&0.12&0.14&-0.047&-0.056&0.069&0.031\nl
log I$_{6724/3727}$&-1.48&-1.44&-1.52&-1.53&-1.27&-1.27\nl
log I$_{1909/5007}$&-1.01&-0.99&-0.91&-0.88&\nodata&-0.78\nl
log I$_{3869/5007}$&-0.72&-0.73&-0.66&-0.67&-0.50&-0.46\nl
\cutinhead{Model Input Parameters\tablenotemark{a}}
T$_{eff}$ (10$^3$K) &&93&&77&&100 \nl
log L/L$_{\sun}$ &&5.2&&3.3&&1.5  \nl
N$_e$ &&30&&30&&30 \nl
R$_o$(pc)&&0.032&&0.032&&0.032 \nl
R (pc)&&5.9\tablenotemark{b}&&1.8&&0.41\tablenotemark{b} \nl
He/H &&0.12&&0.12&&0.12 \nl
O/H ($\times 10^4$)&&5.19&&3.33&&4.45 \nl
C/O &&1.03&&0.72&&0.72 \nl
N/O &&0.71&&0.44&&0.44 \nl
Ne/O &&0.39&&0.32&&0.31 \nl
S/O ($\times 10^3$) &&3.49&&2.63&&3.93 \nl
Ar/O ($\times 10^3$) &&11.1&&11.4&&11.4 \nl
\enddata
\tablenotetext{a}{These parameter values are those necessary to match 
line-of-sight observations only.  They do {\it not} necessarily correspond 
with the real values for the nebula as a whole, simply because 
conditions along one direction usually do not represent global nebular 
conditions.}
\tablenotetext{b}{Matter bounded models}
\end{deluxetable}

\begin{deluxetable}{lccc}
\tablecolumns{4}
\tablewidth{0pc}
\tablenum{4B}
\tablecaption{Correction Factors ($\xi$)}
\tablehead{
\colhead{Ratio} &
\colhead{A} &
\colhead{B} &
\colhead{C}
}
\startdata
He/H&0.96&0.92&0.90 \nl
O/H&0.98&0.92&1.00 \nl
C/O&0.95&0.67&0.56 \nl
N/O&1.14&0.99&0.79 \nl
Ne/O&0.70&0.51&0.31 \nl
S/O&0.76&0.72&0.73 \nl
Ar/O&0.70&0.57&0.56
\enddata
\end{deluxetable}

\begin{deluxetable}{lllllllllll}
\scriptsize
\tablecolumns{11}
\tablewidth{0pc}
\tablenum{5}
\tablecaption{Derived Abundances, Temperatures, \& Densities}
\tablehead{
\colhead{Ratio} &
\colhead{A} &
\colhead{B} &
\colhead{C} &
\colhead{Ave} &
\colhead{PLTP\tablenotemark{a}} &
\colhead{H\tablenotemark{b}} &
\colhead{O\tablenotemark{c}} &
\colhead{KB\tablenotemark{d}} &
\colhead{Orion\tablenotemark{e}} &
\colhead{Sun\tablenotemark{f}}
}
\startdata
He/H&0.12($\pm$0.017)&0.12($\pm$0.018)&0.12($\pm$0.017)&0.12&0.13&0.19&0.14&0.12&0.10&0.10 \nl
O/H($\times 10^{4}$)&4.78($\pm$1.35)&4.36($\pm$1.27)&4.67($\pm$1.29)&4.60($\pm$0.18)&9.12&3.3&4.30&4.78&5.25&7.41 \nl
C/O&0.98($\pm$1.72)&0.75($\pm$0.73)&\nodata&0.87($\pm$0.12)&\nodata&\nodata&\nodata&1.15&0.59&0.48 \nl
N/O&0.74($\pm$0.26)&0.43($\pm$0.17)&0.44($\pm$0.18)&0.54($\pm$0.14)&$>$0.39&0.95&0.89&0.47&0.11&0.13 \nl
Ne/O&0.39($\pm$0.21)&0.32($\pm$0.18)&0.29($\pm$0.21)&0.33($\pm$0.04)&$>$0.63&0.80&\nodata&0.26&0.15&0.16 \nl
S/O($\times 10^{3}$)&3.57($\pm$2.0)&2.94($\pm$1.9)&3.16($\pm$1.7)&3.22($\pm$0.26)&\nodata&\nodata&\nodata&17.4&28.2&28 \nl
Ar/O($\times 10^{3}$)&7.77($\pm$3.5)&6.48($\pm$3.3)&5.97($\pm$2.9)&6.74($\pm$0.76)&8.71&\nodata&\nodata&5.13&5.89&4.47 \nl
T$_{[O~III]}$($\times 
10^{-3}$K)&9.1($\pm$0.88)&9.5($\pm$0.44)&9.4($\pm$0.46)&9.3($\pm$0.17)&10.2&10.7&11.7&\nodata&\nodata&\nodata \nl
T$_{[N~II]}$($\times10^{-3}$K)&9.6($\pm$0.37)&9.7($\pm$0.53)&9.8($\pm$0.38)&9.7($\pm$0.08)&8.2&9.3&9.4&\nodata&\nodata&\nodata 
\nl
T$_{[O~II]}$($\times10^{-3}$K)&15.2($\pm$1.3)&8.3($\pm$0.30)&9.3($\pm$0.40)&10.9($\pm$3.0)&\nodata&\nodata&\nodata&\nodata&\nodata&\nodata 
\nl
N$_e$(cm$^{-3}$)&$<$100&$<$100&$<$100&$<$100&100&570&\nodata&\nodata&\nodata&\nodata \nl
\enddata
\tablenotetext{a}{Helix abundance results from Peimbert, Luridiana, \& Torres-Peimbert (1995)}
\tablenotetext{b}{Helix abundance results from Hawley (1978)}
\tablenotetext{c}{Helix abundance results from O'Dell 1998}
\tablenotetext{d}{Average PN abundances for the complete sample of Kingsburgh \& 
Barlow (1994)}
\tablenotetext{e}{Abundances (gas+dust) for the Orion Nebula from Table~19 of Esteban et al. (1998)}
\tablenotetext{f}{Solar abundances from Grevesse, Noels, \& Sauval (1996)}
\end{deluxetable}

\newpage
\figurenum{1}\figcaption{Logarithmic negative greyscale image of the Helix Nebula in H$\alpha$
with our KPNO (rectangles) and IUE (ovals) slit positions and
orientations illustrated.
The image was taken on the
Palomar 1.5m telescope with a focal-reducing lens system developed by J.
Hester.}
\figurenum{2A}\figcaption{IUE SWP spectrum of position B. Emission
lines of interest are indicated.}
\figurenum{2B}\figcaption{Same as 2B but for LWP.}
\figurenum{2C}\figcaption{Optical spectrum of position~B obtained with the Goldcam spectrograph
and blue grating attached to the KPNO 2.1m telescope. Emission lines of interest
are marked. Lines marked with an asterisk have
components from the night sky, whose effects were difficult to nullify
completely. We do not report strengths for these lines nor do we use
them in any calculations.}
\figurenum{2D}\figcaption{Same as C. but for the red grating.}
\figurenum{3}\figcaption{Density contour map of the Helix Nebula made from our continuum
subtracted H$\alpha$ image, where the field is identical to that of 
Figure~1.  The figure shows iso-density contours normalized
to the average density of the center of the nebula.  Each contour step is in
ratio units of $\pm$0.1 with a range from 0.5 in the outer parts to 1.6 in
the brightest rim area NNE of the central star.}
\figurenum{4}\figcaption{A 3D surface representation of the line-of-sight density, where north is to the 
bottom right and east is to the top right.  Note the elevated
surface brightness of the center compared to the outer perimeter of the
nebula, and relative flatness of that region within the brighter ring
of the nebula.}
\figurenum{5}\figcaption{Calibrated greyscale ratio map of
[N~{\sc ii}]$\lambda$6584 divided by H$\alpha$
for the Helix Nebula with continuum and stars
subtracted.  The greyscale is linear, ranging from
0 (white) to 3 (black) for the ratio values. The left panel presents the full 16\arcmin~ field, while the right panel presents an enlarged view of the area around the norhtern rim of 8\arcmin.   Note the many
``plumes'' emanating from knots, where the plumes are
bright in [N~{\sc ii}] at their heads
(dark on our ratio image).  See text for discussion.}
\figurenum{6}\figcaption{Upper panel: H$\alpha$ surface brightness in 10$^{-15}$ erg
cm$^{-2}$ s$^{-1}$ arcsec$^{-2}$ versus distance from the central star
in arcseconds, where 1$\arcsec=1\times 10^{-3}$pc=3.2$\times
10^{15}$cm, assuming a distance to the nebula of 213pc.  Line type is
used to designate position angle of the sample cut, as defined in the
legend. These data have been deprojected, as described in the text.
The  bold lines represent model results described in the text.  Values
from the spectrophometric observations reported in {\S}2.2 are
indicated with X symbols.  Lower panel:~same as upper panel but for the
intensity ratio [N~II] $\lambda$6584/H$\alpha$.}
\figurenum{7}\figcaption{A. Predicted intensity ratios of He~II
$\lambda$4686/H$\beta$ versus distance from the central star in
arcseconds for the model simulations of nebular conditions along
position angles 135 and 315 degrees (solid line) and 24 and 209
degrees (dashed line). Values from the spectrophometric
observations reported in {\S}2.2 are indicated with X symbols. 
Spectrophotometric measurements from O'Dell (1998) are shown with short 
horizontal lines and error bars, where the latter indicate the length along the 
slit of the extracted spectrum.
B. Same as A. but for [O~II]
$\lambda$3727/H$\beta$. C.~Same as A. but for [O~III]
$\lambda$5007/H$\beta$.}
\figurenum{8}\figcaption{A. Fractional ionization of the three ions of helium as a
function of distance from the central star in arcseconds.  The solid
line shows the model results for the position angles 135$^{\circ}$ and 315$^{\circ}$, while the dashed line shows the results for position angles 24$^{\circ}$
and 209$^{\circ}$. B. Same as a. but for fractional ionization of oxygen.
C. Same as A. but for electron temperature. X symbols give values for
[O~III] electron temperatures derived in {\S}3.3. D.  Same as A. but
for electron density. X symbols give values for [S~II] electron
densities derived in {\S}3.3. E. Same as A. but for the ratio of disk
thickness (Z) to diameter (D).}

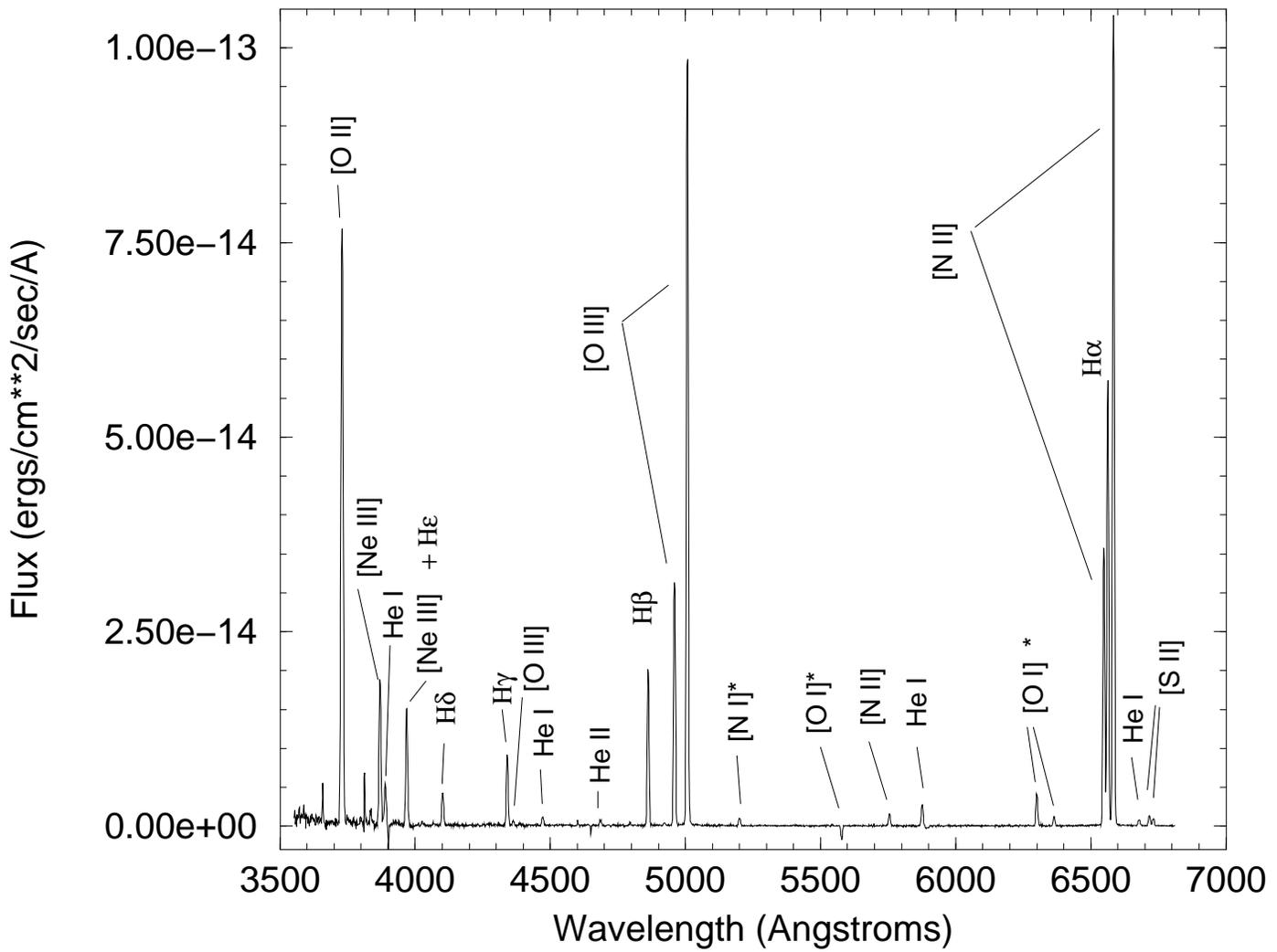
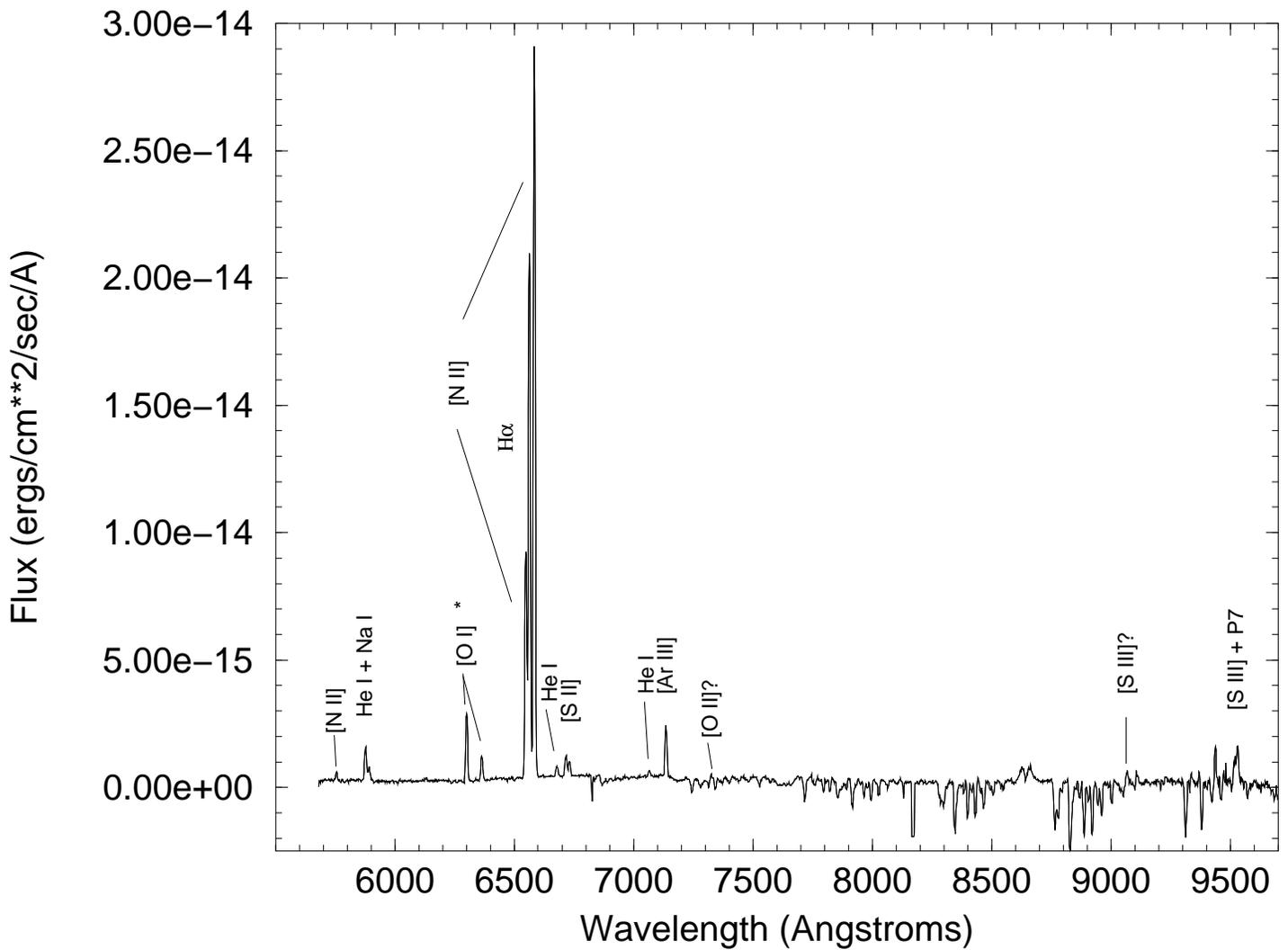
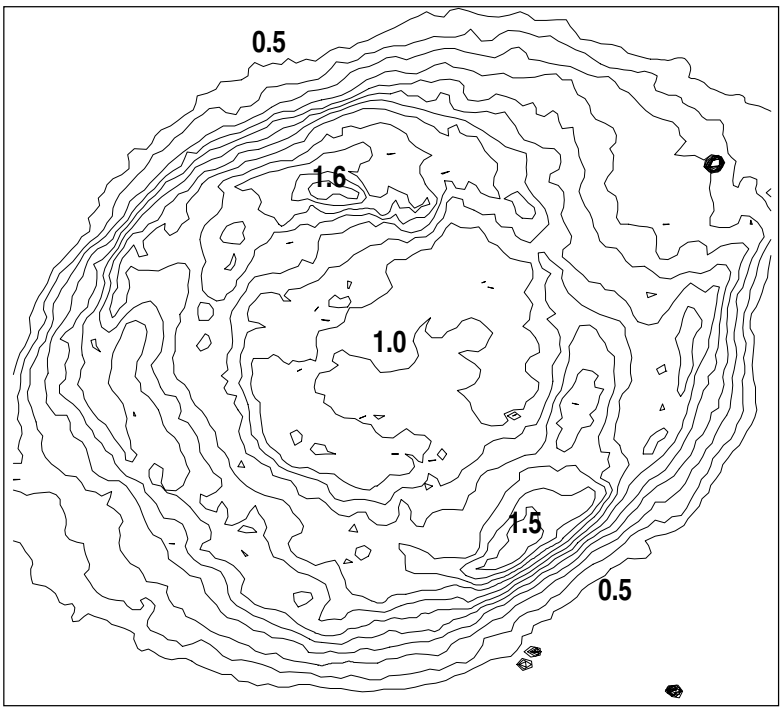
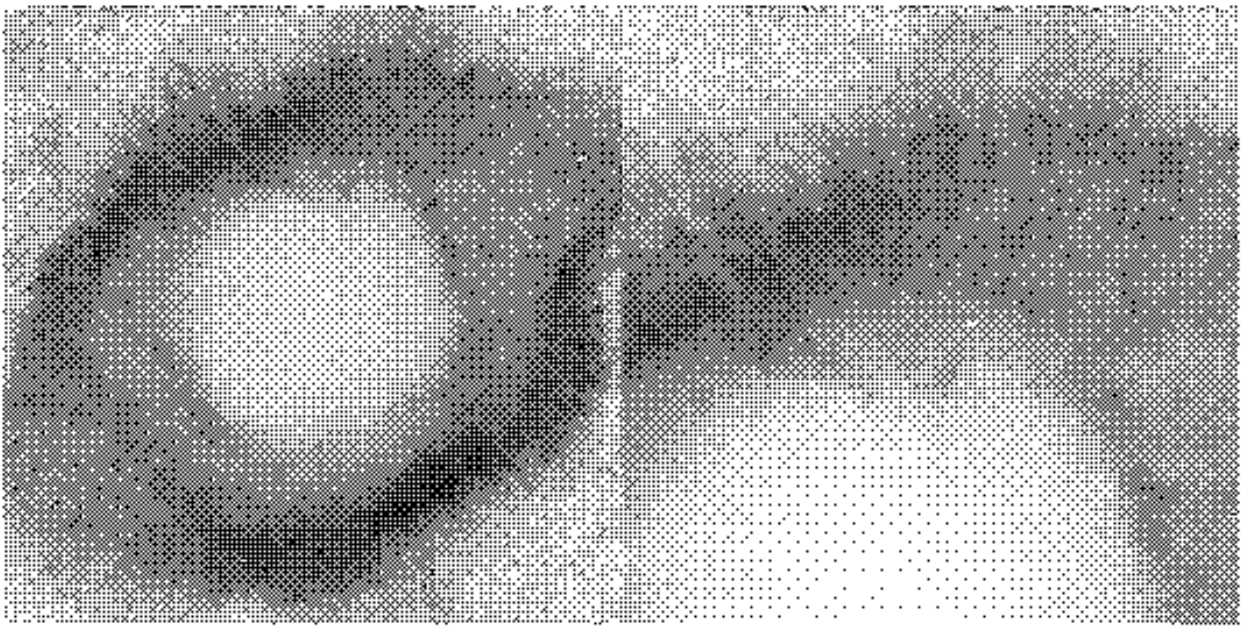
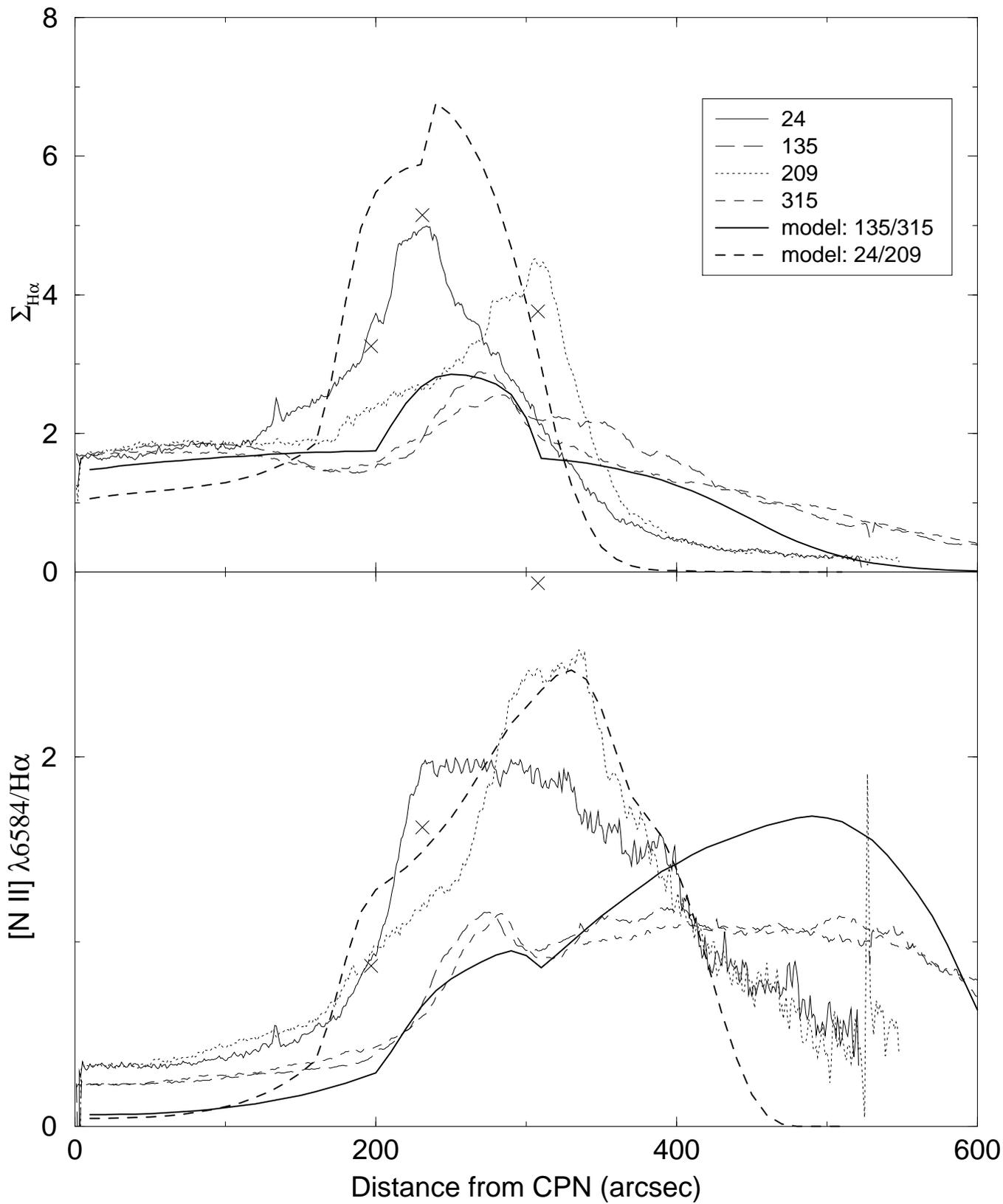
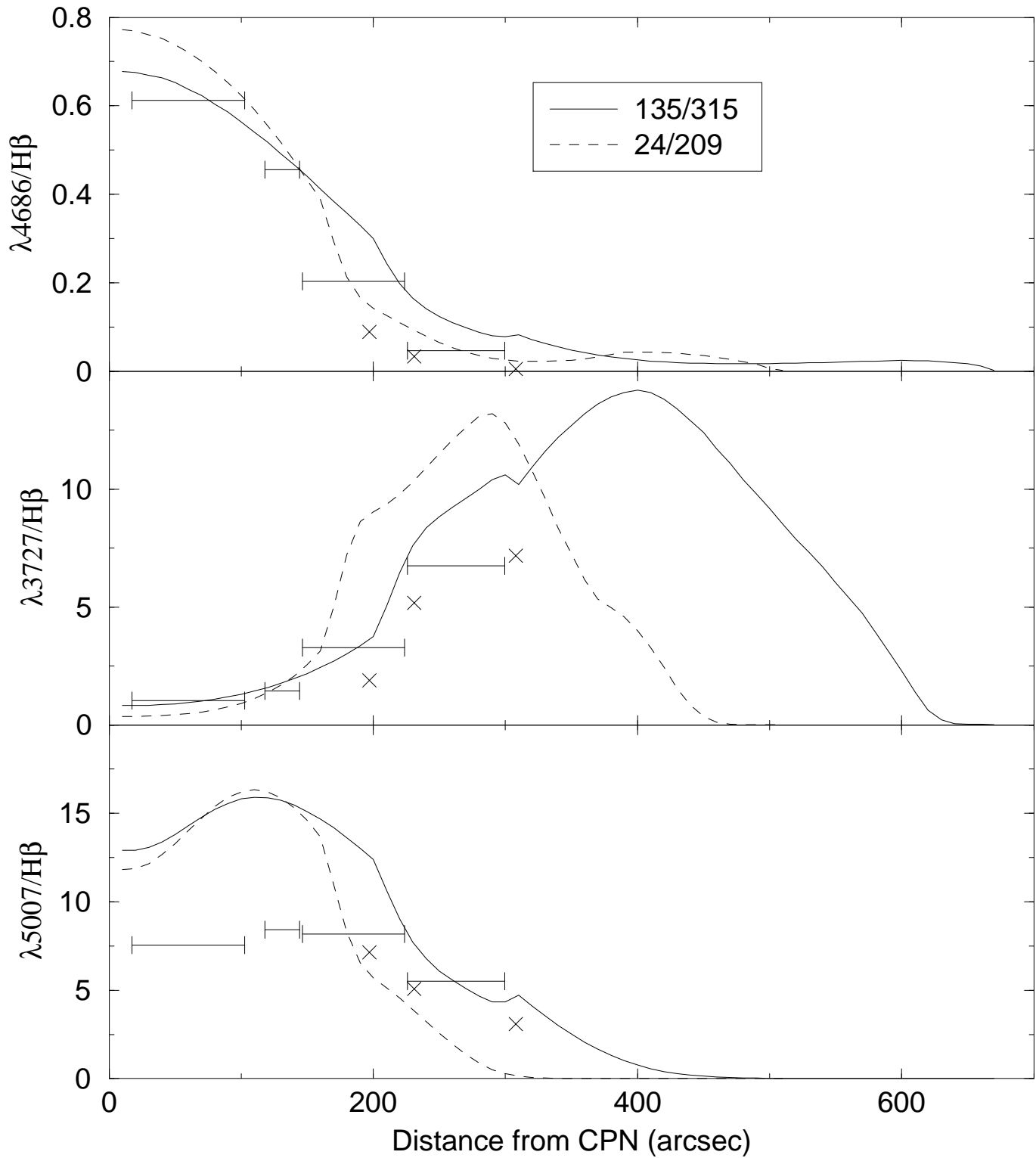
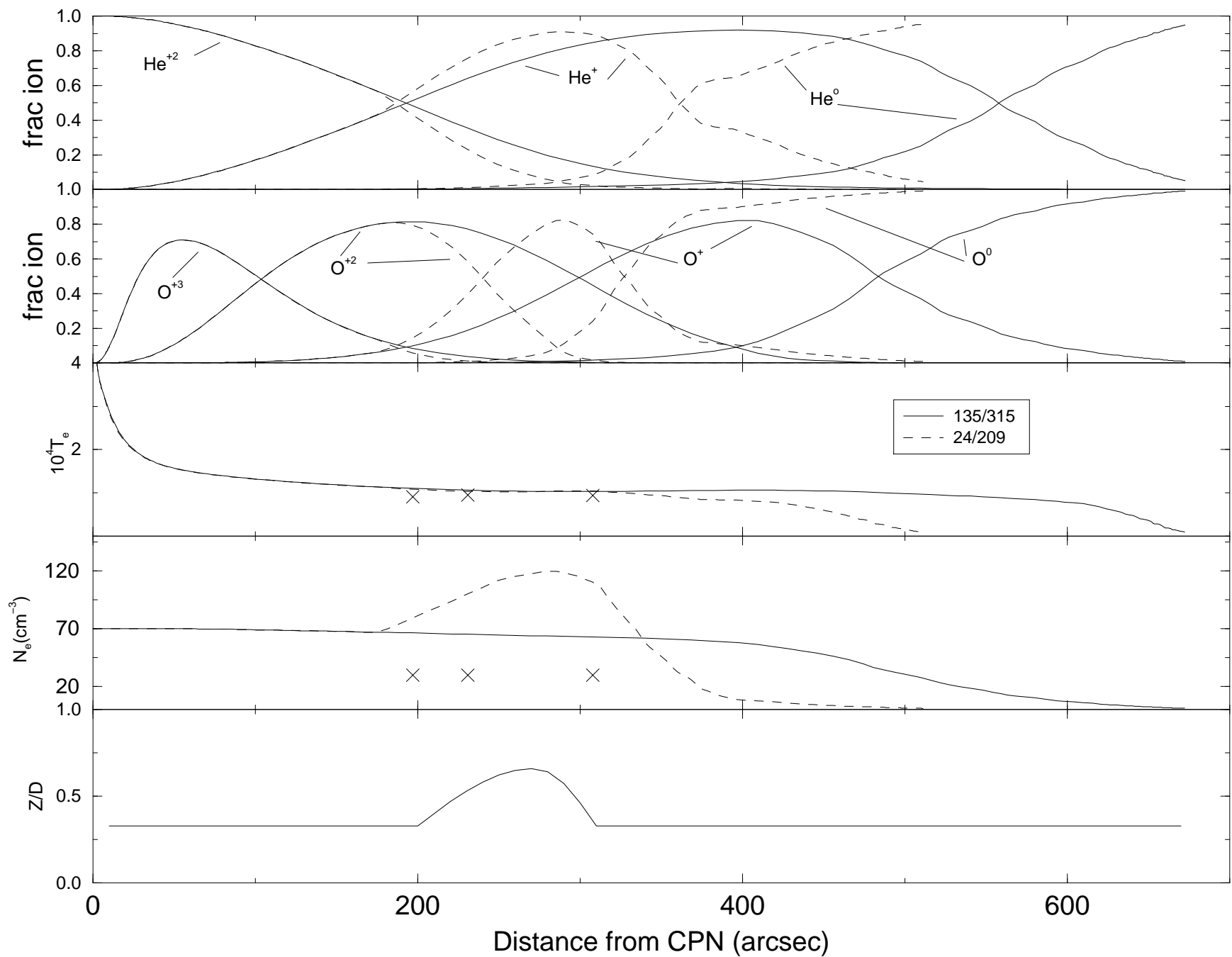
\end{document}